\documentclass[11pt]{article}

\usepackage[final]{acl}

\usepackage{times}
\usepackage{latexsym}

\usepackage[T1]{fontenc}

\usepackage[utf8]{inputenc}

\usepackage{microtype}

\usepackage{inconsolata}

\usepackage{graphicx}

\usepackage{hyperref}
\usepackage{url}
\usepackage{booktabs}
\usepackage{amsmath,amssymb,bm}
\usepackage{epsfig,subfigure,caption}
\usepackage{algpseudocode}
\usepackage[normalem]{ulem}
\usepackage[linesnumbered,algoruled,boxed,noend]{algorithm2e}
\usepackage{xcolor}
\usepackage{color, colortbl}
\usepackage{multirow, hhline}
\usepackage{wrapfig}
\usepackage{listings}
\usepackage{adjustbox}
\usepackage{makecell}
\usepackage{lineno} 
\usepackage{dsfont}

\newcommand{\hitl}{\texttt{HitL}\xspace}
\newcommand{\llama}{\texttt{Llama-3-70B}\xspace}

\definecolor{darkblue}{rgb}{0, 0, 0.5}
\hypersetup{colorlinks=true, citecolor=darkblue, linkcolor=darkblue, urlcolor=darkblue}

\title{Uncovering Intervention Opportunities for Suicide Prevention with Language Model Assistants}

\author{
 \textbf{Jaspreet Ranjit\textsuperscript{1}},
 \textbf{Hyundong J. Cho\textsuperscript{2}},
 \textbf{Claire J. Smerdon\textsuperscript{1}},
 \textbf{Yoonsoo Nam\textsuperscript{1}},
\\
 \textbf{Myles Phung\textsuperscript{1}},
 \textbf{Jonathan May\textsuperscript{2}},
 \textbf{John R. Blosnich\textsuperscript{3}},
 \textbf{Swabha Swayamdipta\textsuperscript{1}}
\\
\\
 \textsuperscript{1}Thomas Lord Dept. of Computer Science, University of Southern California,\\
 \textsuperscript{2}Information Sciences Institute, University of Southern California,\\
 \textsuperscript{3}Suzanne-Dwork School of Social Work, University of Southern California
\\
 \small{
   \textbf{Correspondence:} \href{mailto:jranjit@usc.edu}{jranjit@usc.edu}
 }
}
\begin{document}

\maketitle
\begin{abstract}

The National Violent Death Reporting System (NVDRS) documents suicides in the United States.
In a demanding public health data pipeline, annotators manually extract structured information from death investigation records
following extensive codebooks (i.e. annotation guidelines) painstakingly developed by experts.
In this work, we facilitate data-driven insights from the NVDRS data to support the development of novel suicide interventions by leveraging language models (LM) as assistants to these (a) data annotators and (b) experts. 
We find that LM predictions match existing data annotations about 85\% of the time across 50 NVDRS variables. 
Where the LM disagrees with existing annotations, our expert review identifies that 38\% of these instances reveal inconsistencies between narratives and structured data. 
Finally, we introduce a human-in-the-loop algorithm that helps experts efficiently build and refine codebooks for new variables by having them only focus on providing feedback for incorrect LM predictions. 
We apply our algorithm to a real-world case study, and find that about 96K narratives contain evidence of victim interactions with legal professionals, which surfaces a substantial opportunity for upstream intervention that is not captured in the original structured data. 
Our findings provide evidence that LMs can serve as effective assistants to public health researchers who handle sensitive data in high-stakes scenarios.

\end{abstract}

\section{Introduction}
\begin{figure*}[h!]
  \includegraphics[width=\textwidth]{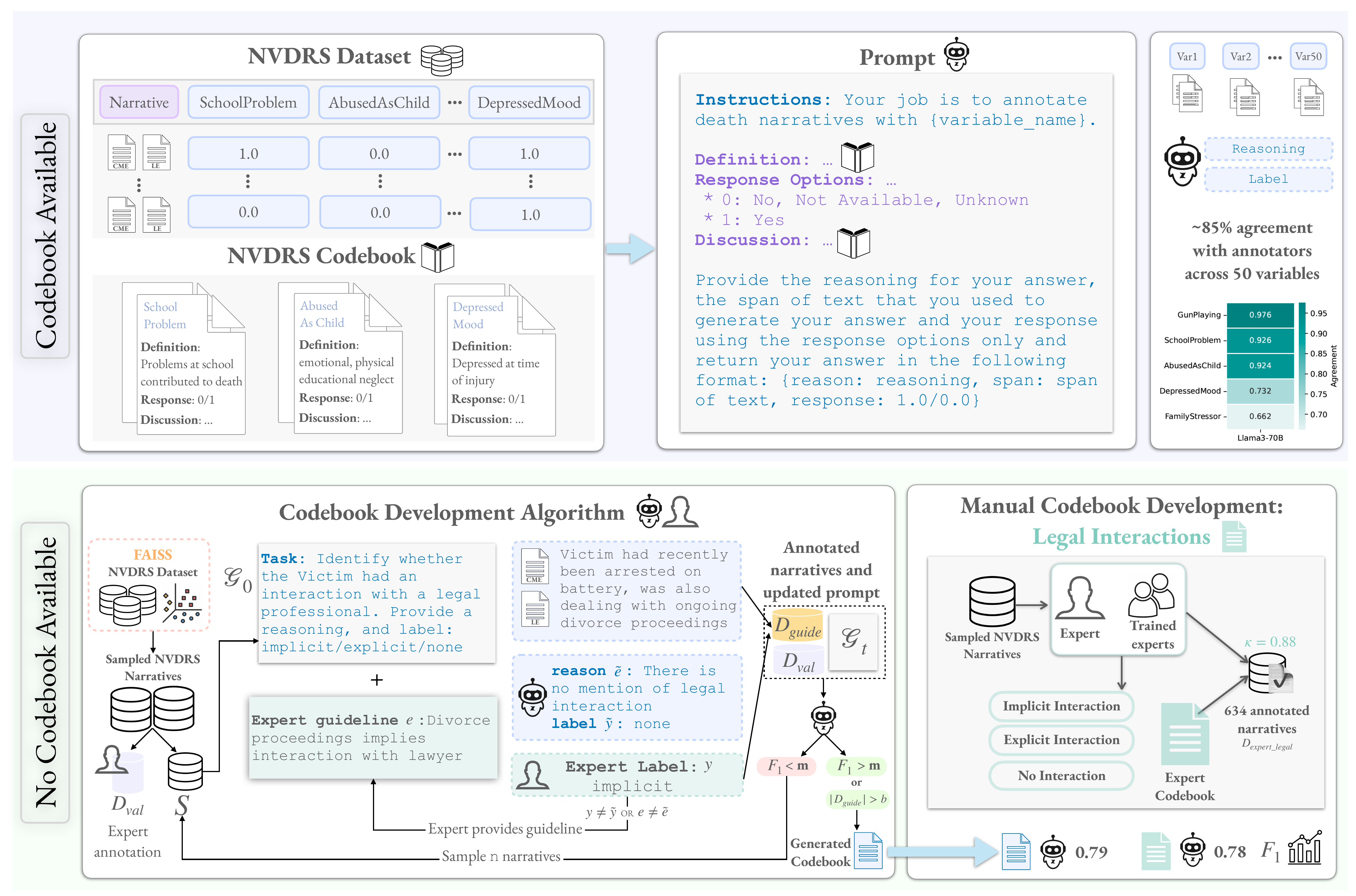}
  \caption{
  LM assistants can help annotate NVDRS narratives when a codebook is available (top) and help experts efficiently develop new codebooks for novel variables (bottom). 
  When a codebook is available, we incorporate it in the prompt and generate predictions. 
  For a new variable of interest, we propose a human-in-the-loop codebook development algorithm.
  Here, experts focus on providing feedback for incorrect LM predictions, reducing manual codebook development time from weeks to hours (bottom right). 
  We apply our algorithm to a novel variable characterizing victim interactions with legal professionals. We find that 10.4\% of 270K NVDRS narratives contain evidence of such interactions, indicating a new and salient intervention opportunity, while achieving annotation quality comparable to a laborious manual approach.  
 }

  \label{fig:nvdrs_pitch_fig}
  \vspace{-0.5cm}
\end{figure*}

{\textcolor{red}{{Warning: This paper discusses suicide, which may be distressing to readers.}}}

Each year, approximately 50,000 people in the United States fall victim to suicide~\citep{cammack2024vital}. 
The Centers for Disease Control and Prevention (CDC) documents this information in the National Violent Death Reporting System (NVDRS)\footnote{\url{https://www.cdc.gov/nvdrs/about/index.html}}, which contains structured and unstructured data for more than 270K suicides (\autoref{fig:nvdrs_pitch_fig}; top left).
Structured data in NVDRS is manually labeled by annotators (or NVDRS data abstractors\footnote{We use the term data abstractors and annotators interchangeably.}) using codebooks---precise definitions and annotation guidelines---developed painstakingly by experts.
Given the sensitive nature of the topic and the scale of NVDRS, a data annotator's job is demanding and emotionally taxing~\citep{fincham2008impact, murthy2024national, nazarov2019research}. 
Moreover, variation in manual reporting and data abstraction leaves room for annotation inconsistencies \citep{dang2023research}. 

Given its scale, NVDRS is a valuable source for data-driven discovery of \textit{novel} suicide interventions\footnote{Identifying and measuring novel variables linked to suicide risk that can serve as evidence to motivate the more effective deployment of existing interventions.}.
To uncover actionable insights for new suicide interventions, experts must move beyond existing structured variables and extract evidence from unstructured narratives that support their hypotheses. 
However, this process requires experts to manually analyze NVDRS narratives, develop a codebook for new variables, and abstractors to retroactively annotate all 270K cases. 

To overcome these challenges, we investigate whether language models (LMs) can help public health researchers make data annotation and analysis for suicide prevention research more efficient and effective.
Our investigation is based on the promise LMs have shown in social science \citep{rytting2023towards, pangakis2023automated, halterman2024codebook, ziems2024can}, and aims to answer two research questions. 

\textbf{RQ1:} \textit{Can a language model alleviate the burden of annotating variables with existing reference codebooks in NVDRS and surface discrepancies between the structured and unstructured data?} 
For \textbf{RQ1}, we deploy LMs as annotation assistants to data abstractors and find that they achieve an average agreement of 85\% with data abstractors across 50 variables (\textsection \ref{sec:llm_validation}), making them a reliable peer validator for high-agreement variables. 
They are also useful for low-agreement variables: from our expert review of six low agreement variables, we find that 38\% of disagreement instances reveal inconsistencies between the narratives and the structured data, suggesting that LMs can surface inconsistencies between the two sources of data (\textsection \ref{sec:llm_inconsistency}).

\textbf{RQ2:} \textit{Can a language model assist experts in the annotation of new variables that go beyond current NVDRS codebooks, potentially leading to new intervention opportunities?} 
For \textbf{RQ2}, we introduce a human-in-the-loop codebook development algorithm that helps experts iteratively develop codebooks for new variables by providing feedback for incorrect LM predictions (\textsection \ref{sec:codebook-development}). 
We first demonstrate the effectiveness of our algorithm by developing codebooks for existing variables using only the variable name as our starting point.
Our experiments show that our algorithm produces codebooks that enable LMs to achieve an average 80\% agreement with data annotators, outperforming LMs conditioned on the existing official NVDRS codebooks (75\% agreement) for 12 variables (\textsection \ref{sec:agent-in-the-loop}). 

Suicide prevention research has largely focused on risk factors arising in clinical settings, leaving opportunities for \textit{preventative} care through interactions with nonclinical professionals underexplored~\cite{lybarger2023advancements, ralevski2024using, xu2024analyzing}. 
To examine this gap, we apply our algorithm to study a novel nonclinical factor for suicide prevention: identifying victim interactions with legal professionals (\textsection \ref{sec:case_study}). 
We find that 12.7\% of NVDRS narratives contain evidence of such interactions, indicating a substantial and underexplored avenue for new interventions.
Furthermore, our algorithm reduces codebook development time from weeks to hours without compromising on annotation quality, demonstrating its potential to efficiently surface and characterize a wider range of nonclinical intervention opportunities (e.g., interactions with other nonclinical professionals).

In summary, our results indicate the ability of LMs to serve as efficient assistants that enable experts to mobilize the NVDRS dataset more effectively, and accelerate the identification of novel intervention opportunities\footnote{Our code is publicly available: \url{https://dill-lab.github.io/interventions_lm_assistants/}}. 
The promising results we demonstrate in building LM assistants for professionals in public health emphasize the value of collaborative approaches between LMs and human experts, especially in high-stakes domains.

\section{Predicting Predefined Variables in NVDRS Narratives}
\label{sec:llm_annotations}

We describe the NVDRS narratives and structured data (\textsection \ref{sec:nvdrs}), and address our first research question (\textsection \ref{sec:annotating_nvdrs_intro}): Can LMs alleviate the burden of manual annotation for NVDRS narratives? 
This setting includes established codebooks and abstractor labels, allowing us to evaluate whether LM predictions can follow the same. 

\subsection{The National Violent Death Reporting System (NVDRS)}
\label{sec:nvdrs}

The NVDRS contains records for 270K suicide cases between 2003-2019 in 42 US states, the District of Columbia, and Puerto Rico \citep{liu2023surveillance, wilson2022surveillance}.
Each case is characterized by more than 600 structured variables which include victim demographics, and circumstances surrounding their death \citep{liu2023surveillance}. 
Suicide circumstance (e.g., eviction or loss of home, alcohol problem) and crisis (e.g., events occurring within two weeks of death) variables are annotated using four data sources: death certificates, coroner / medical examiner (CME) records, law enforcement (LE) records, and crime laboratory records. 
Narratives are recorded using information from the CME / LE records \citep{paulozzi2004cdc}. 
Data abstractors annotate cases by following an extensive codebook containing annotation guidelines for each variable.
In NVDRS, variables are used to identify populations at elevated risk (e.g., intimate partner violence, housing instability) and to inform the development, and evaluation of prevention programs (e.g., lethality assessments, referral pathways, training initiatives for frontline professionals)~\citep{teutsch2000principles}.
 
\subsection{Experimental Setup}
\label{sec:annotating_nvdrs_intro}

We consider a subset of variables that have been manually annotated by NVDRS abstractors, based on information from CME / LE records.
Since the narratives are also derived from these records \citep{paulozzi2004cdc}, they should ideally capture the same information as the structured data.\footnote{
We selected 50 binary variables (36 circumstance and 14 crisis variables) that appeared in at least 300 cases to ensure reliable evaluation.}

We use locally-hosted, open-weight LMs from various model families and sizes as annotation assistants, given the private and sensitive nature of the NVDRS records. 
We curate LM prompts for each variable by adapting the NVDRS codebooks in a standardized format with instructions, definitions, response options, and discussion, as shown in \autoref{fig:nvdrs_pitch_fig} (top panel). 
For the input, we concatenate the CME and LE narratives, and generate Chain-of-Thought reasoning (CoT; \cite{wei2022chain}), and a label in a zero-shot setting. 
Our approach offers a lightweight alternative to finetuning models~\citep{wang2023nlp}: it is easily adaptable to codebook changes and does not require retraining models, which is computationally expensive. 

For evaluation, we consider a balanced dataset ($|D_\text{balanced}|=$ 25{,}000) where we sample 500 narratives per variable with equal representation across 0/1 classes, to address the high class imbalance for each variable. 
We also evaluate on a random sample of narratives ($|D_\text{random}|=$ 1,000), which more closely reflects real-world deployment conditions where class distributions are heavily skewed. 
In both settings, we report agreement\footnote{Agreement is computed as the ratio of matching labels to total number of instances.} with data abstractor annotations per variable.

\begin{table}[t]
\centering
\begin{adjustbox}{width=0.9\columnwidth}
\begin{tabular}{lccc}
\toprule
\textbf{Model} & \textbf{Mean Agreement} & \textbf{95\% CI} & \textbf{SD} \\
\midrule
Llama-3-70B & \textbf{0.85} & \textbf{[0.82, 0.87]} & 0.09 \\
Qwen2.5-14B & 0.79 & [0.76, 0.82] & 0.10 \\
Qwen2.5-7B & 0.73 & [0.70, 0.75] & 0.08 \\
Mistral-7B & 0.73 & [0.71, 0.75] & 0.07 \\
Llama-3-8B & 0.71 & [0.69, 0.74] & 0.09 \\
\bottomrule
\end{tabular}
\end{adjustbox}
\caption{Mean agreement with data abstractor annotations, 95\% confidence intervals (bootstrap, 10K iters), and standard deviation across 50 variables (SD). \llama achieves the highest agreement. 
Performance is reported on a balanced evaluation set of 500 narratives per variable ($D_\text{balanced}$).}
\label{tab:balanced_eval}
\vspace{-0.5cm}
\end{table}


\begin{figure*}[t]
  \includegraphics[width=\textwidth]{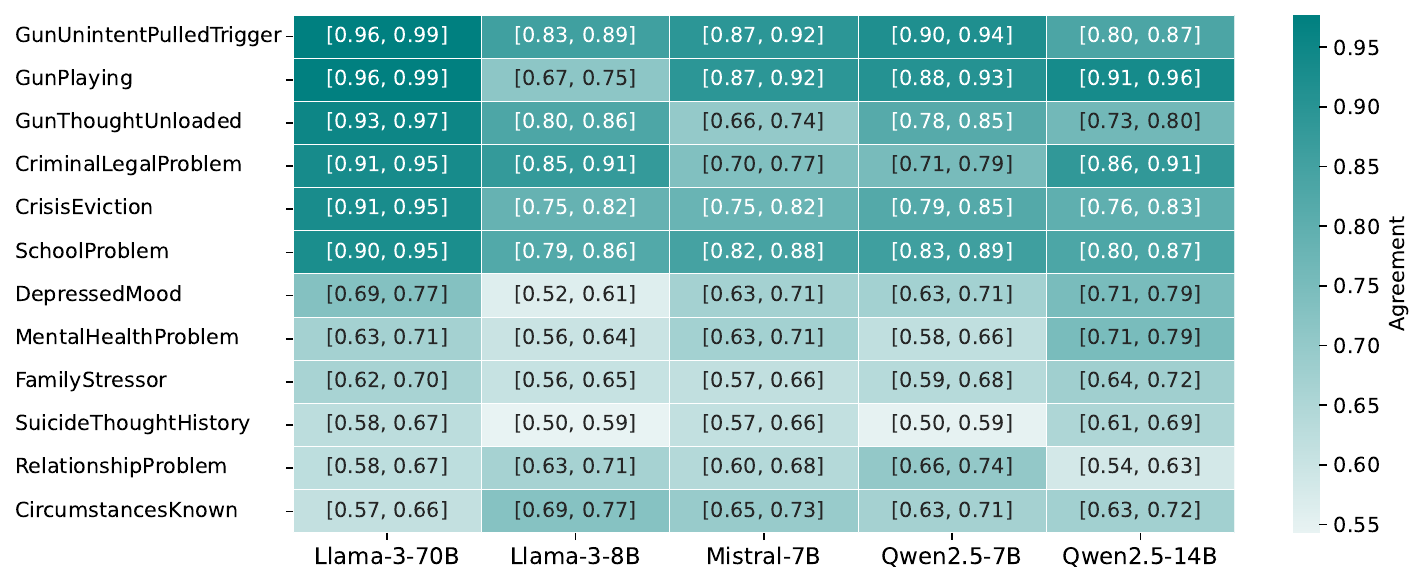}
\caption{Per variable agreement (95\% confidence intervals; bootstrapped with 10K iterations) for 12 highest and lowest agreement variables across all models. \llama has low agreement with data abstractors for mental health-related variables (e.g., {\fontfamily{cmss}\selectfont SuicideThoughtHistory, MentalHealthProblem}) and relatively higher agreement for firearm-related variables (e.g., {\fontfamily{cmss}\selectfont GunPlaying, GunUnintentPulledTrigger}) (See \autoref{fig:per_variable} in Appendix for results across all variables).  
  }
  \vspace{-1em}
  \label{fig:per_variable_condense}
\end{figure*}

\subsection{LM Annotation Results and Analysis}
\label{sec:llm_validation}

\autoref{tab:balanced_eval} shows the average agreement between LM predictions and abstractor annotations for $D_\text{balanced}$ across 50 variables, with 95\% confidence intervals. 
The complete results on $D_\text{random}$ are included in \autoref{tab:random_eval} in Appendix \ref{sec:app_agreement}.
We find that \llama achieves the highest average agreement: 85\% for $D_\text{balanced}$ and 82\% for $D_\text{random}$.
We performed paired t-tests comparing \llama against each model across 50 variables. 
All comparisons were statistically significant ($p<0.0125$ under Bonferroni correction). 
Additionally, \llama outperforms all other models on at least 38 out of 50 variables.
\llama is also highly self-consistent across three runs with varying temperature values (i.e. 0.2, 0.5, 0.7) producing identical predictions 99\% of the time.
Given the fairly high agreement with abstractors, our findings suggest that LM assistants can serve as a peer validator in the absence of additional human verification for NVDRS annotations, especially given only 5\% of annotations are validated by two annotators \citep{liu2023surveillance}.
We qualitatively observe that the LM's CoTs highlight relevant context in long narratives that abstractors may overlook during initial annotation. 
For example, for the {\fontfamily{cmss}\selectfont FamilyStressor} variable, the CoT connects domestic violence history and concerns about the daughter’s suicidal ideation as indicators of a stressful family environment.

From analyzing the 6 highest and 6 lowest agreement variables from $D_\text{balanced}$ (illustrated in \autoref{fig:per_variable_condense}), we observe that all models have relatively higher agreement for firearm-related variables (e.g., 
{\fontfamily{cmss}\selectfont GunPlaying, GunUnintentPulledTrigger}) and low agreement for mental health-related variables (e.g., {\fontfamily{cmss}\selectfont SuicideThoughtHistory, MentalHealthProblem}). 
The true positive rate (TPR) and false negative rate (FNR), as shown in \autoref{tab:tpr_fnr}, also indicate that concrete, observable events with explicit lexical cues (e.g., {\fontfamily{cmss}\selectfont GunUnintentPulledTrigger}) achieve very high TPR ($>$0.96) and low FNR ($<$0.03). 
In contrast, variables related to mental health and emotional state (e.g., {\fontfamily{cmss}\selectfont MentalHealthProblem, DepressedMood, SuicideThoughtHistory}) exhibit low TPRs and high FNRs ($>$0.7). 

To understand why LMs fall short in some cases for low agreement variables, we conducted a targeted qualitative analysis for two of them: {\fontfamily{cmss}\selectfont MentalHealthProblem} and {\fontfamily{cmss}\selectfont DepressedMood}. 

We randomly sampled 100 \llama predictions from $D_\text{balanced}$, including 25 false negatives and 25 false positives for each variable (MentalHealthProblem and DepressedMood), and manually analyzed their chains-of-thought (CoT) to identify two recurring failure modes. 
For each case, we manually reviewed the narratives and LM prediction using a rubric that assessed: (i) whether relevant evidence was present in the narrative, (ii) whether evidence was explicit vs. implicit, and (iii) whether the model’s CoT cited appropriate evidence to make the prediction. 
Across both variables, we observe that 88\% of false negative cases contained relevant evidence in only one of the two concatenated narratives. In contrast, 70\% of false positive cases had CoTs that cited circumstances (e.g., financial stress, substance use, interpersonal conflict) as evidence for mental health problems, even though these factors are not valid indicators of mental health problems according to the NVDRS annotation guidelines. 

\subsection{LM Disagreements Surface Annotation Inconsistencies}
\label{sec:llm_inconsistency}

Given variations in recording practices, prior work by \citet{wang2023nlp, wang2024natural} reveal inconsistently annotated NVDRS variables among abstractors. 
Specifically, \citet{wang2024natural} identify cases where circumstances described in the narratives are not reflected in the structured data, resulting in inconsistencies between the two sources.
Given their finding, we hypothesize that disagreement between the LM assistant and the abstractor can surface inconsistencies between the narratives and structured data.
For six variables with low agreement between the LM (\llama) and abstractor (i.e.
{\fontfamily{cmss}\selectfont CircumstancesKnown, RelationshipProblem, FamilyStressor, MentalHealthProblem, DepressedMood, HistoryMentalIllnessTreatment}),
we sample 150 narratives where the LM and abstractor disagree and another 150 where they agree. 
For each of these settings, we asked for a second opinion from a suicide prevention expert on our team with 15 years of experience who was asked to independently annotate 300 cases without knowledge of the LM or abstractor annotations. 

In cases where the LM and the abstractor disagree, the expert finds that the original annotation is inconsistent with the information contained in the narrative 38\% of the time, compared to only 13\% of the time when the LM and the abstractor are in agreement. 
Given all variables are binary, we define an annotation inconsistency as either: (i) the narrative contains evidence for the variable but was labeled 0 by the abstractor, or (ii) the narrative lacks evidence for the variable but was labeled 1. 
We conclude from a bootstrap hypothesis test for equality of means \citep{efron1994introduction} that this difference is statistically significant ($p < 0.05$), indicating the potential for our assistant to surface annotation inconsistencies between the narratives and structured data (see Appendix \ref{sec:app_inconsistencies} for discussion on the robustness of our hypothesis test).
Future work is needed to systematically account for and rectify annotation inconsistencies in the NVDRS annotations across a larger number of variables with LM assistants.

\section{Characterizing New Variables in NVDRS Narratives}
\label{sec:new-variables}

Although NVDRS codebooks have continuously expanded their scope, they are not exhaustive and updates are infrequent and incremental due to the scale and diversity of narratives \citep{steenkamp2006national}.
Therefore, it is paramount to develop efficient methods for characterizing new variables to enable data-driven analysis for testing novel hypotheses in prevention research and informing intervention strategies~\citep{blair2016national}. 

To this end, we introduce an algorithm that efficiently creates codebooks for new variables by refining an initially coarse codebook. 
In this section, we elaborate on our algorithm (\textsection \ref{sec:codebook-development}) and demonstrate its usefulness by validating LM-developed codebooks through simulations with existing variables (\textsection \ref{sec:agent-in-the-loop}). 
In the following section, we present a case study to extract real-world, actionable insights for a new variable capturing victims' interactions with legal professionals (\textsection \ref{sec:case_study}).

\subsection{Codebook Development Algorithm for Characterizing Variables}
\label{sec:codebook-development}

Our algorithm, formally defined in Algorithm \ref{alg:our_algorithm}, 
uses an LM to (i) generate predictions using the current codebook in the same manner as \textsection \ref{sec:llm_annotations}, and (ii) iteratively refine the codebook by synthesizing per-sample feedback from experts on those LM predictions.\footnote{We use \llama with different instructions for both generating predictions and refining the codebook.} 
The main advantage of our algorithm over a traditional manual codebook development (\textsection \ref{sec:expert_guideline_development}) is that the expert only needs to focus on providing feedback on LM predictions for a few samples in each iteration, making it more efficient. 
The codebook is automatically refined by the same LM with a codebook update prompt based on the expert's feedback. 
We repeat this process until the codebook results in LM predictions that achieve a pre-defined target performance on an evaluation set.\looseness=-1

\RestyleAlgo{ruled}
\SetKwComment{Comment}{/* }{ */}
\algnewcommand{\algorithmicand}{\textbf{ and }}
\algnewcommand{\algorithmicor}{\textbf{ or }}
\algnewcommand{\AND}{\algorithmicand}
\algnewcommand{\OR}{\algorithmicor}

\begin{algorithm}[t]
\small
\caption{
    Codebook Development 
}
\label{alg:our_algorithm}
\KwIn{$\pi_\theta$: LM, $\mathcal{D}=\{x_i\mid 1 \leq i \leq N\}$: full set of unannotated NVDRS narratives,  $\mathcal{D}_{guide}^{0}=\emptyset$: human-validated NVDRS narratives, $\mathcal{D}_{val}$: human-annotated NVDRS narratives, $\mathcal{G}_0$: initial guideline, $\mathcal{U}$: guideline update prompt, $m$: target accuracy, $k$: minimum $\mathcal{D}_{guide}$ set size, $b$: budget, $t$: iteration index, $\mathcal{F}$: feedback}
\KwOut{$\mathcal{D}_{\text{LM}}$: LM-annotated data}, $\mathcal{G}_t$: final guideline

$t \gets 0 $ \textcolor{gray}{// Iteration 0}

\While{
True
}{

$\mathcal{S} \sim \mathcal{D} \,  \backslash \, \mathcal{D}_{guide}^{t-1}, |S| = k$ \hfill \textcolor{gray}{// Sample from $\mathcal{D}$}

$\mathcal{I} \gets \emptyset $

\For{$x \in \mathcal{S}$}{
    $\tilde{y}, \tilde{e} \sim \pi_{\theta}(\mathcal{G}_t(x))$
    \hfill \textcolor{gray}{// Generate LM label and reasoning}

    $y, e \sim \mathcal{F}(x, \tilde{y}, \tilde{e})$ \hfill \textcolor{gray}{// Feedback provides correct label and reasoning}

    \If{$\tilde{y}\neq y \OR \tilde{e} \neq e $}{
    $\mathcal{I} = \mathcal{I} \cup \{(x, \tilde{y}, y, \tilde{e}, e)\} $
\hfill \textcolor{gray}{// Collect LM errors}
    
    }

    $\mathcal{D}_{guide}^{t} = \mathcal{D}_{guide}^{t-1} \cup \{(x, y, e\}$
}

$\mathcal{G}_{t+1} \gets \pi_\theta(\mathcal{U}(\mathcal{G}_t, I))$ \hfill \textcolor{gray}{// Update guideline based on LM errors}

$ t \gets t+1 $

\If{$\mathcal{D}_{guide}^{t} \neq \emptyset  $}{
    $acc \gets \frac{\sum_{(x, y, e) \in \mathcal{D}_{val}} \mathds{1}[\pi_{\theta}(\mathcal{G}_t(x)) = y]}{|\mathcal{D}_{val}|}$
\hfill \textcolor{gray}{// Check for stopping criteria}

    \If{$acc \geq m \And |\mathcal{D}_{guide}^{t}| \geq k$ \OR $|\mathcal{D}_{guide}^{t}| > b$}{

    \textbf{break}
    }
}

}

$\mathcal{D}_{\text{LM}} \gets \{(x, y, e) \mid 
y, e \sim \pi_{\theta}(\mathcal{G}_t(x)), x \in \mathcal{D} \, \backslash \, \mathcal{D}_{{guide}}^{t}\}$ \hfill \textcolor{gray}{// Annotate $\mathcal{D}$ with final $\mathcal{G}$}

\Return $\mathcal{D}_{\text{LM}}$, $\mathcal{G}_t$
\end{algorithm}

\paragraph{Initialization.}
First, we collect expert annotations on a subset of narratives, $\mathcal{D}_{val}$, which serves as a held-out validation set, containing at least $j$ instances per class.
The likelihood of finding a true positive can be low depending on the expert hypothesis.
To address this, we initialize $\mathcal{D}$ by upsampling instances based on expert-defined keywords relevant to the variable of interest, using similarity search with FAISS \citep{douze2024faiss}. 
We set our initial codebook $\mathcal{G}_0$ using a template that only includes the variable name and label classes (see \autoref{tab:prompt_templates} in Appendix \ref{sec:app_case_study}).
The LM uses this to predict a label $\tilde{y}$ along with its chain-of-thought reasoning $\tilde{e}$ for $n$ sampled narratives at each iteration $t$. \looseness=-1

\paragraph{Sampling Example Narratives}
We consider random and coverage-based sampling~\citep{gupta-etal-2023-coverage} to select $n$ narratives to annotate in each iteration.
The latter selects $n$ narratives that are most dissimilar to those seen in prior iterations ($\mathcal{D}_{guide}^{t}$) such that experts avoid redundant cases. 
We detail our coverage-based sampling in Appendix \ref{sec:coverage-based-sampling}.

\paragraph{Codebook update with feedback.}
For each incorrect LM prediction, our expert provides the corrected label $y$, and a free-text rationale $e$ explaining their label.
The current $\mathcal{G}_t$ is then updated by prompting an additional language model to incorporate $e$ into $\mathcal{G}_{t+1}$.
Each expert-validated annotation is added to the growing evaluation set $\mathcal{D}_{guide}^{t}$ and the LM's performance is evaluated using $\mathcal{G}_t$ on $\mathcal{D}_{guide}^{t}$ and $\mathcal{D}_{val}$ upon each update to $\mathcal{G}_t$.
This process is repeated until performance on $\mathcal{D}_{val}$ exceeds a specified target performance $m$, or the number of expert-validated annotations in $\mathcal{D}_{guide}^{t}$ exceeds a predefined budget $b$. 
The budget refers to the maximum number of samples the expert can annotate in the duration of the algorithm.
We set a minimum size $k$ for $|\mathcal{D}_{guide}^t|$ to account for the target $m$ being met prematurely. 
A simplified overview of the process is shown in the bottom of \autoref{fig:nvdrs_pitch_fig}. 

\begin{figure*}[h!]
\centering
  \includegraphics[width=0.9\textwidth]{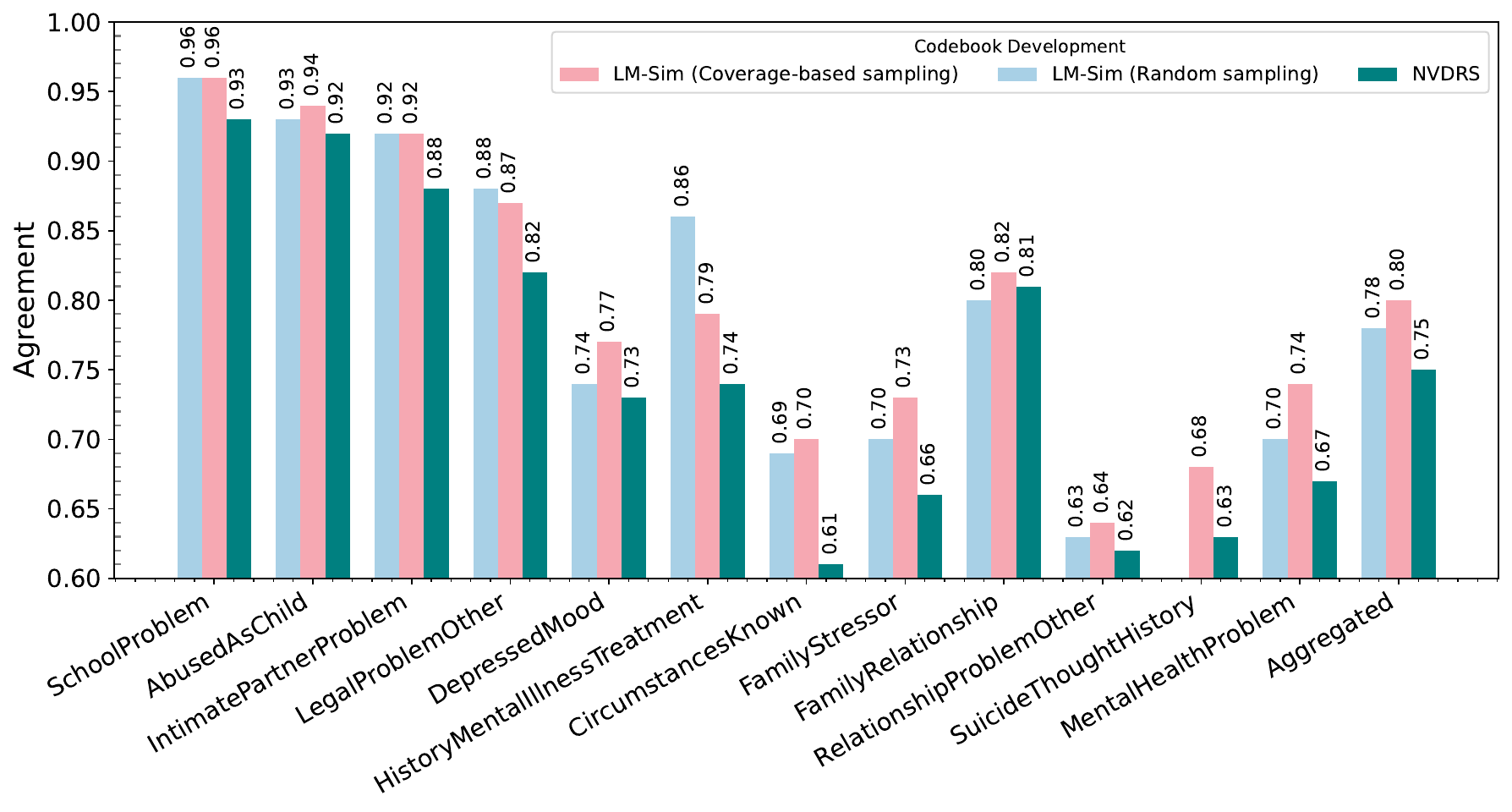}
  \vspace{-0.5cm}
\caption{\llama performance on $D_\text{balanced}$ using the generated codebooks (LM-Sim) for random and coverage-based sampling on 12 variables, compared to performance achieved using the reference NVDRS codebook. Our generated codebooks from the simulated setting are just as effective as using the NVDRS codebook for all 12 variables as shown by the `Aggregated' agreement ($\sim$0.80 for LM-Sim vs $\sim$0.75 for NVDRS).
  }
  \label{fig:nvdrs_hitl_acc_sim}
  \vspace{-0.5cm}
\end{figure*}

\subsection{Simulated Codebook Development: Existing NVDRS Variables}
\label{sec:agent-in-the-loop}

\paragraph{Simulation Setup.}
To validate the usefulness and generalizability of our codebook development algorithm before we apply it to real-world new variables, we first evaluate it on a subset of existing variables by treating them as new variables that are not yet characterized using simulated expert feedback from an LM. 
To scale our experiments across multiple NVDRS variables, we use existing NVDRS abstractor labels $y$ (see \textsection \ref{sec:llm_annotations}) as reference labels
and LM-based expert feedback ($e$) as a proxy for human feedback.  
While we have reference $y$, we do not have corresponding reference reasoning, $e$. 
To address this, we simulate an expert reasoning via the LM's CoT (as generated in \textsection \ref{sec:llm_annotations}) 
for $e$ (refer to further details in Appendix \ref{sec:simulated_codebook_development_details}). 
We apply our algorithm in this simulated setting to generate codebooks for 12 variables, using a subsample $\mathcal{D}$ of 150 narratives per variable, balanced across 0/1 classes.
The 12 variables are chosen to cover varying degrees of average LM-annotator agreement as reported in \textsection \ref{sec:llm_annotations} (four each from low (0.6-0.7), medium (0.7-0.8) and high agreement (0.8-0.9)) so that we can evaluate how well the generated codebooks perform across different levels of annotation difficulty. 
In our simulated setting, the algorithm terminates after reaching the annotation budget ($b$=150 narratives). 
We include our hyperparameter settings in Appendix \ref{sec:app_hyperparameters}. 

We measure the effectiveness of the generated codebooks by measuring their corresponding LM prediction performance using the same prompt template and held out test set $D_\text{balanced}$ from \textsection \ref{sec:llm_annotations} and compare
them to the performance achieved using the reference NVDRS codebooks. 

\paragraph{Simulation Results.}
Paired t-tests ($p<0.025$ under Bonferroni correction) across 12 variables show that our generated codebooks significantly outperform the NVDRS codebook for random and coverage-based sampling as shown in \autoref{fig:nvdrs_hitl_acc_sim}. 
Furthermore, the maximum accuracy on $\mathcal{D}_{guide}^{t}$ is reached between 10-15 iterations, as shown in \autoref{fig:llm_sim_all_var} (see \autoref{fig:llm_sim_all_var_coverage} in Appendix \ref{sec:app_hyperparameters} for results on all variables). 
We observe that our generated codebooks provide finer-grained instructions to the model by including examples from the narratives, which helps resolve ambiguities that may be underspecified in NVDRS codebooks (see \autoref{tab:codebook_comps} in the Appendix \ref{sec:app_codebooks_comp}).

Our simulation results provide preliminary evidence that our algorithm enables efficient collaboration between LMs and experts that result in effective codebooks. 
However, simulation results are not sufficient to validate the usefulness of our algorithm. 
In the following section, we apply our algorithm for a real-world case study to derive a codebook that is used to collect data-driven evidence for uncovering new intervention opportunities.\looseness=-1

\section{
Case Study: Legal Interactions of Suicide Victims}
\label{sec:case_study}

Suicide prevention experts have identified legal professionals as part of a broader set of nonclinical `industries of disruption' (e.g. financial advisors, homeless shelters) that frequently interact with at-risk individuals. 
Yet, prevention efforts largely focus on the clinical sector~\citep{labouliere2018zero, wang2023nlp, consoli2024sdoh, guevara2024large}, which primarily engages in \textit{remedial} care, leaving opportunities for \textit{preventive} care through interactions with nonclinical professions underexplored~\citep{lybarger2023advancements, ralevski2024using, xu2024analyzing}. 
A recent survey found that 40\% of attorneys had a client die by suicide, 70\% had concerns about a client’s suicide risk, but 65\% had never received suicide prevention training~\citep{blosnich2024industries}. 
However, NVDRS lacks structured variables that capture these interactions, preventing a direct analysis which could inform new preventive measures (e.g., training lawyers to spot risk factors in clients). 
For our case study, our goal is to measure the prevalence of nonclinical (legal) interactions and use this evidence to motivate how existing suicide prevention efforts can be more effectively applied in nonclinical settings, where such efforts may not currently exist. For example, our findings can support the development of gatekeeper training for legal professionals who frequently interact with at-risk individuals.

Identifying interactions with legal professionals is challenging because they are often indirectly implied by references to life disruptions (e.g. custody battles), where there is a lot of ambiguity (e.g., DUIs may not always involve legal interactions).
Existing methods based on retrieval methods \citep{kafka2023detecting, Kafka125} cannot capture these implicit instances.
Therefore, we characterize interactions with legal professionals into three classes: explicit (direct interactions), implicit (indirect interactions inferred from life disruptions), and no interaction.
Since feedback comes from a human expert for our algorithm in a real-world case study setting, we refer to our algorithm-based codebook development as human-in-the-loop (\hitl) codebook development.

\subsection{\hitl Codebook Development}
\label{sec:hitl_guideline_development}

We apply our algorithm to NVDRS narratives to answer how many cases contain evidence of victim interactions with legal professionals, to determine whether such nonclinical contexts are relevant for prevention.
Given a budget of 150 instances, and batch size of 5, we first sample narratives using FAISS~\citep{douze2024faiss} with the following keywords: `\texttt{lawyer, attorney}.' 
We start with a simple codebook ($\mathcal{G}_{\text{0}}$) as shown in \autoref{fig:nvdrs_pitch_fig}. 
In our pilot, the human expert exhausts the entire budget $b$.\looseness=-1

\subsection{Manual Codebook Development}
\label{sec:expert_guideline_development}
We conduct this case study with a manual approach that does not leverage human-LM collaboration to create a held-out test set and assess the efficiency of our algorithm. 
Based on the expertise of the suicide prevention researcher on our team (who is a co-author of this work), 
we manually develop codebooks to identify and characterize victims' interactions with legal professionals~\citep{halterman2024codebook, rytting2023towards}.
Our manual codebook development involved three annotators: one suicide prevention expert and two trained CS graduate students
All three independently annotated 150 narratives over two weeks. This was followed by a 2-hour discussion to refine definitions and resolve disagreements for achieving consensus. This process was repeated twice on two additional sets of 100 narratives each, with each round taking approximately one week of independent annotation followed by a 1 hour discussion, until no further refinements to the definitions were needed. 
These guidelines are then used by all three to annotate 634 narratives ($D_{\text{expert\_legal}}$), as an evaluation set for our algorithm. 
On this set, we achieve high inter-annotator agreement:  Krippendorff's $\alpha=0.88$ \citep{krippendorff1970estimating}. 


\begin{table}[t]
\centering
\small
\begin{tabular}{lccc}
\toprule
\textbf{Model} & \textbf{$\mathcal{G}_{0}$}  & \textbf{$\mathcal{G}_{\hitl}$} & \textbf{$\mathcal{G}_{\text{expert}}$} \\
\midrule
Llama-3-70B & 0.57 & 0.80 & 0.78 \\
Qwen2.5-32B & 0.68 & 0.76 & 0.77  \\
Qwen2.5-14B & 0.63 & 0.75 & 0.77 \\ 
\bottomrule
\end{tabular}
\caption{Macro F$_1$ for implicit-, explicit-, and no-interaction reported on $D_{\text{expert\_legal}}$ using no codebook ($\mathcal{G}_{0}$), the \hitl generated codebook at the 25th iteration ($\mathcal{G}_{\hitl}$), and the expert codebook ($\mathcal{G}_{\text{expert}}$). The generated codebook achieves performance on par with the expert codebook across all models.
}
\vspace{-1.5em}
\label{tab:hitl_lawyer}
\end{table}

\subsection{\hitl Results on Legal Interactions}
\label{sec:evaluating_hitl}

In \autoref{tab:hitl_lawyer}, we compare performance on $D_{\text{expert\_legal}}$ across three prompts: (\textit{i}) \textbf{$\mathcal{G}_{\text{0}}$}: No Codebook, (\textit{ii}) \textbf{$\mathcal{G}_{\text{expert}}$}: Expert Codebook (\textsection \ref{sec:expert_guideline_development}), and (\textit{iii}) \textbf{$\mathcal{G}_{\text{\hitl}}$}: \hitl Codebook at the 25th iteration.
We observe that $\mathcal{G}_{\text{\hitl}}$ performance (Macro F$_1$) is on par with $\mathcal{G}_{\text{expert}}$ across all models suggesting that our guidelines can be generalized beyond the model that was used to develop them.
Our \hitl codebook development pilot was more efficient, requiring 3.5 hours of expert time and 125 annotated narratives by focusing feedback on a smaller set of informative cases, compared to approximately four weeks and 350 narratives in the manual setting. 
Additionally, our results show that we do not compromise on annotation quality for victim-legal interactions. 
Most importantly, our algorithm finds that \textbf{12.7\% of all 756K\footnote{This expanded set includes 483,303 deaths (2003–2023) and 756,645 narratives, with 87,005 cases having a single C/ME or LE narrative and 334,820 having both.} suicide narratives in NVDRS contain evidence of legal interactions with victims}.
More concretely, \llama,
prompted using our \hitl codebook, detects implicit victim-lawyer interactions in 12.16\% of the narratives and explicit interactions in 0.53\% of the narratives (see \autoref{fig:interaction_distribution} in the Appendix \ref{sec:app_case_study}). 
These findings support the efficiency and effectiveness of our algorithm for experts to verify novel hypotheses, such as victim interactions with additional nonclinical stakeholders (e.g. financial institutions, housing shelters) \citep{sinyor2024effect}. 
Most importantly, our finding that interactions between suicide victims and legal professionals are relatively common, may provide the required empirical evidence for designing novel interventions in suicide prevention, such as training for legal professionals.
We emphasize that such efforts should be pursued through collaboration with domain experts, where LMs should augment (e.g. via validation) rather than replace human expertise.

\section{Related Work}
\label{sec:related_work}

\paragraph{Characterizing NVDRS Data using NLP.} 
Despite the scale of NVDRS narratives, relatively few studies have examined the effectiveness of LMs for characterizing them \citep{dang2023research}. 
Most prior work has focused on using AI tools to identify risk factors such as social determinants of health using electronic health record (EHR) and medical notes~\citep{kirtley2022translating, lejeune2022artificial, ehtemam2024role, melia2025application, johns2023understanding, zhou2023identifying, consoli2024sdoh, guevara2024large, wang2025multi}, which are limited to individuals who have engaged with healthcare systems.  
For NVDRS narratives, NLP tools have been used to characterize these narratives by narrative length and quality~\cite{arseniev2023gendered} and variation across decedents’ demographics~\cite{chance2025measuring}. 
Others uncover latent themes~\citep{arseniev2022integrating, arseniev2021aggression, davidson2021exploring} or 
classify circumstance and crisis variables from narratives~\cite{wang2023nlp, zhou2023identifying, kafka2023detecting, parker2025supervised}.
Prior work typically frame LMs as annotation assistants, without enabling the discovery of \textit{new} intervention opportunities. 
Motivated by these limitations, we examine whether LMs can serve as effective assistants to data abstractors and experts (see Appendix \ref{sec:app_relatedwork} for further related work). 

\paragraph{Qualitative Coding with LMs.}
Prior work has explored the use of language models for thematic analysis \citep{katz2024thematic, dai2023llm}, qualitative coding \citep{barany2024chatgpt, tai2024examination, xiao2023supporting, perkins2024use}, and annotation of socially sensitive data \citep{ranjit-etal-2024-oath, halterman2024codebook, rytting2023towards, pangakis2023automated}. 
Despite these advances, it remains unclear how expert judgment can be effectively integrated into LM-assisted codebook development for real-world, high-stakes domains \citep{wang2024large, ziems2024can}. 
In particular, prior work has not examined LM-assisted codebook refinement in the context of NVDRS, which presents unique challenges due to the sensitivity and heterogeneity of the data, which to our knowledge our work is the first attempt towards.

\section{Conclusion}

We introduce LM assistants to help human experts efficiently develop codebooks and human abstractors to accurately annotate structured NVDRS variables. 
Our LM assistant achieves high agreement with NVDRS abstractors while surfacing annotation inconsistencies, and our algorithm produces codebooks that result in annotation performance on par with using the reference codebooks. 
We further apply our algorithm to a real-world case study on a new variable: interactions between suicide victims and legal professionals.
Our findings about the high prevalence of such interactions could open intervention avenues in suicide prevention research in future work. 
Recognizing the potential risks of using LMs in high-stakes domains, we analyze failure modes to inform responsible practitioner use and advise against relying on LMs as peer validators for low-agreement variables, particularly those related to mental health.
Also, models should flag high uncertainty predictions or abstain altogether, prioritizing safety over coverage. 
Finally, codebook development should always be supervised by domain experts as LM-only approaches pose serious safety risks in high-stakes contexts, thus the success of such systems depend on effective human-LM collaboration, where LMs augment rather than replace human expertise.

\section*{Limitations}
\label{sec:limitations}

We show that LM assistants can support the large-scale validation of sensitive death narrative annotations, and enable the discovery of data-driven evidence for new intervention opportunities. 
However, given the scope, real-world implications, and unique challenges of the suicide prevention domain, we acknowledge some important limitations. 

First, the evaluation of our real-world case study relies on 634 expert-annotated narratives, which required ~3 months to annotate. Obtaining ground-truth validation for all model predictions across 270K narratives was infeasible at this rate, given the emotionally demanding nature of the task.

Another limitation of our work is the scope of our empirical evaluation---our real-world case study intentionally focuses on a single hypothesis within suicide prevention: victim interactions with legal professionals.
A larger range of nonclinical factors may also be relevant for prevention. 
More broadly, while our codebook development algorithm is not domain specific, evaluating its effectiveness beyond the suicide prevention context was outside the scope of this work. 

Additionally, while our codebook development algorithm supports iterative guideline development, it does not incorporate multi-turn interaction between experts and LMs in an effort to minimize expert burden. However, we recognize that this design choice may have limited opportunities to clarify codebooks for ambiguous cases. 

We do not explicitly assess whether codebook development achieves thematic saturation: the point at which the codebooks sufficiently capture all relevant cases observed for a given variable~\citep{groundingtheory}. In practice, assessing thematic saturation is challenging as it would require additional human-in-the-loop experiments across many hyperparameter settings, which was not feasible for our case study. 
In the future, we recommend practitioners take into account the subjectivity of the variable and the manual effort involved when determining the stopping conditions and hyperparameters (i.e., budget, target performance and the minimum size of $\mathcal{D}_{guide}^t$ and $\mathcal{D}_{val}$), which affect how quickly thematic saturation is achieved.

Our evaluation in \textsection \ref{sec:llm_validation} is limited to a subset of 50 binary NVDRS variables that occur in at least 300 cases, and that can be inferred from the narrative data which we have access to. As a result, we do not evaluate the full set of 600+ NVDRS variables which are derived from external data sources such as death certificates or toxicology reports which we do not have access to. This constraint may limit the applicability of our findings to only those variables that were evaluated in our study.

\section*{Ethics Statement}
Our work explores the use of language models to facilitate data-driven insights from suicide death narratives in the National Violent Death Reporting System (NVDRS). 
This project was reviewed by our Institutional Review Board, who deemed it as NOT human subjects research; therefore approval was not necessary. 
In the following sections, we outline ethical considerations, including annotation procedures, annotator well-being, model limitations, and data privacy protections.

\paragraph{Annotator Well-Being.}
Given the sensitive nature of suicide death narratives, we implemented several precautions in line with current best practices~\cite{dasgupta2021ghost} to support annotator well-being. 
The supervising expert engaged with the study team on a bi-weekly basis to have debriefs and check-ins regarding any feelings of emotional or mental heaviness of the task. Aside from formal team meetings, the supervising expert was also available to all study team members for individual appointments to discuss any concerns, emotional reactions, or mental strains from annotating. Specifically, the supervising expert instructed the annotators to be mindful of their reactions and feelings while annotating and, if they felt even the slightest inclination to stop annotating, then they should stop and engage in some activity that they enjoy (e.g., exercise, watch television, be with friends, etc.). 

It is important to emphasize that NVDRS data \textit{rarely} contain suicide notes~\cite{pestian2012sentiment, rockett2018discerning}; in some scant instances, only paraphrases may appear.
While the NVDRS narratives, themselves, may be graphic, researchers who have worked with both suicide notes and coroner reports indicate that suicide notes are emotionally heavier pieces of data to process than coroner reports~\cite{fincham2008impact}. 

\paragraph{LMs as Annotation Assistants and Intended Usage.}
We conducted human annotation with one expert in suicide prevention and two trained annotators. 
The expert, a practitioner in the field, provided training, ongoing supervision, and led the codebook development process in collaboration with the annotators. 
All annotators independently labeled the same set of narratives, achieving strong inter-annotator agreement (Krippendorff's $\alpha$ = 0.88). Disagreements were resolved through consensus discussions.
Given the sensitive nature of suicide death narratives, we implemented multiple safeguards to support annotator well-being (see previous paragraph). While we explore the potential of language models to assist with expert annotation, we do not advocate for their deployment as a replacement for human annotators, but as tools to support more efficient and informed expert analysis.

\paragraph{Privacy.}
The NVDRS data is already de-identified for protecting the privacy of the victims and their survivors.
We followed NVDRS protocols for responsible data handling, and all experiments were conducted locally with open-weight models, ensuring that no data was shared with any LM API providers. 
To further protect data privacy, we do not include any qualitative narrative excerpts in the paper and all narratives were de-identified.

\bibliography{custom}

@misc{halterman2024codebook,
  title={Codebook LLMs: Adapting Political Science Codebooks for LLM Use and Adapting LLMs to Follow Codebooks},
  author={Halterman, Andrew and Keith, Katherine A},
  year={2024},
  note={arXiv},
}

@misc{rytting2023towards,
  title={Towards coding social science datasets with language models},
  author={Rytting, Christopher Michael and Sorensen, Taylor and Argyle, Lisa and others},
  note={arXiv:2306.02177},
  year={2023}
}

@article{pangakis2023automated,
  title={Automated annotation with generative ai requires validation},
  author={Pangakis, Nicholas and Wolken, Samuel and Fasching, Neil},
  journal={arXiv preprint arXiv:2306.00176},
  year={2023}
}

@article{kafka2023detecting,
  title={Detecting intimate partner violence circumstance for suicide: development and validation of a tool using natural language processing and supervised machine learning in the National Violent Death Reporting System},
  author={Kafka, Julie M and Fliss, Mike D and Trangenstein, Pamela J and Reyes, Luz McNaughton and Pence, Brian W and Moracco, Kathryn E},
  journal={Injury prevention},
  volume={29},
  number={2},
  pages={134--141},
  year={2023},
  publisher={BMJ Publishing Group Ltd}
}

@article{labouliere2018zero,
  title={“Zero Suicide”--A model for reducing suicide in United States behavioral healthcare},
  author={Labouliere, Christa D and Vasan, Prabu and Kramer, Anni and Brown, Greg and Green, Kelly and Rahman, Mahfuza and Kammer, Jamie and Finnerty, Molly and Stanley, Barbara},
  journal={Suicidologi},
  volume={23},
  number={1},
  pages={22},
  year={2018},
  publisher={NIH Public Access}
}

@article{murthy2024national,
  title={National Strategy for Suicide Prevention},
  author={Murthy, Vivek},
  year={2024}
}

@article{nazarov2019research,
  title={Research utility of the National Violent Death Reporting System: a scoping review},
  author={Nazarov, Oybek and Guan, Joseph and Chihuri, Stanford and Li, Guohua},
  journal={Injury epidemiology},
  volume={6},
  pages={1--12},
  year={2019},
  publisher={Springer}
}

@article{wei2022chain,
  title={Chain-of-thought prompting elicits reasoning in large language models},
  author={Wei, Jason and Wang, Xuezhi and Schuurmans, Dale and Bosma, Maarten and Xia, Fei and Chi, Ed and Le, Quoc V and Zhou, Denny and others},
  journal={Advances in neural information processing systems},
  volume={35},
  pages={24824--24837},
  year={2022}
}

@article{wang2024natural,
  title={A natural language processing approach to detect inconsistencies in death investigation notes attributing suicide circumstances},
  author={Wang, Song and Zhou, Yiliang and Han, Ziqiang and Tao, Cui and Xiao, Yunyu and Ding, Ying and Ghosh, Joydeep and Peng, Yifan},
  journal={Communications Medicine},
  volume={4},
  number={1},
  pages={199},
  year={2024},
  publisher={Nature Publishing Group UK London}
}

@article{wang2023nlp,
  title={An NLP approach to identify SDoH-related circumstance and suicide crisis from death investigation narratives},
  author={Wang, Song and Dang, Yifang and Sun, Zhaoyi and Ding, Ying and Pathak, Jyotishman and Tao, Cui and Xiao, Yunyu and Peng, Yifan},
  journal={Journal of the American Medical Informatics Association},
  volume={30},
  number={8},
  pages={1408--1417},
  year={2023},
  publisher={Oxford University Press}
}

@article{consoli2024sdoh,
  title={SDoH-GPT: Using Large Language Models to Extract Social Determinants of Health (SDoH)},
  author={Consoli, Bernardo and Wu, Xizhi and Wang, Song and Zhao, Xinyu and Wang, Yanshan and Rousseau, Justin and Hartvigsen, Tom and Shen, Li and Wu, Huanmei and Peng, Yifan and others},
  journal={arXiv preprint arXiv:2407.17126},
  year={2024}
}

@article{guevara2024large,
  title={Large language models to identify social determinants of health in electronic health records},
  author={Guevara, Marco and Chen, Shan and Thomas, Spencer and Chaunzwa, Tafadzwa L and Franco, Idalid and Kann, Benjamin H and Moningi, Shalini and Qian, Jack M and Goldstein, Madeleine and Harper, Susan and others},
  journal={NPJ digital medicine},
  volume={7},
  number={1},
  pages={6},
  year={2024},
  publisher={Nature Publishing Group UK London}
}

@misc{lybarger2023advancements,
  title={Advancements in extracting social determinants of health information from narrative text},
  author={Lybarger, Kevin and Bear Don’t Walk IV, Oliver J and Yetisgen, Meliha and Uzuner, {\"O}zlem},
  journal={Journal of the American Medical Informatics Association},
  volume={30},
  number={8},
  pages={1363--1366},
  year={2023},
  publisher={Oxford University Press}
}

@article{ralevski2024using,
  title={Using Large Language Models to Annotate Complex Cases of Social Determinants of Health in Longitudinal Clinical Records},
  author={Ralevski, Alexandra and Taiyab, Nadaa and Nossal, Michael and Mico, Lindsay and Piekos, Samantha N and Hadlock, Jennifer},
  journal={medRxiv},
  year={2024},
  publisher={Cold Spring Harbor Laboratory Preprints}
}

@inproceedings{xu2024analyzing,
  title={Analyzing Social Factors to Enhance Suicide Prevention Across Population Groups},
  author={Xu, Richard Li and Wang, Song and Wang, Zewei and Zhang, Yuhan and Xiao, Yunyu and Pathak, Jyotishman and Hodge, David and Leng, Yan and Watkins, S Craig and Ding, Ying and others},
  booktitle={2024 IEEE 12th International Conference on Healthcare Informatics (ICHI)},
  pages={189--199},
  year={2024},
  organization={IEEE}
}

@article{dang2023research,
  title={Research utility and limitations of textual data in the National Violent Death Reporting System: a scoping review and recommendations},
  author={Dang, Linh N and Kahsay, Eskira T and James, LaTeesa N and Johns, Lily J and Rios, Isabella E and Mezuk, Briana},
  journal={Injury epidemiology},
  volume={10},
  number={1},
  pages={23},
  year={2023},
  publisher={Springer}
}

@article{johns2023understanding,
  title={Understanding suicide over the life course using data science tools within a triangulation framework},
  author={Johns, Lily and Zhong, Chuwen and Mezuk, Briana},
  journal={Journal of psychiatry and brain science},
  volume={8},
  number={1},
  pages={e230003},
  year={2023}
}

@article{zhou2023identifying,
  title={Identifying rare circumstances preceding female firearm suicides: validating a large language model approach},
  author={Zhou, Weipeng and Prater, Laura C and Goldstein, Evan V and Mooney, Stephen J and others},
  journal={JMIR mental health},
  volume={10},
  number={1},
  pages={e49359},
  year={2023},
  publisher={JMIR Publications Inc., Toronto, Canada}
}

@article{arseniev2022integrating,
  title={Integrating topic modeling and word embedding to characterize violent deaths},
  author={Arseniev-Koehler, Alina and Cochran, Susan D and Mays, Vickie M and Chang, Kai-Wei and Foster, Jacob G},
  journal={Proceedings of the National Academy of Sciences},
  volume={119},
  number={10},
  pages={e2108801119},
  year={2022},
  publisher={National Academy of Sciences}
}

@article{arseniev2021aggression,
  title={Aggression, escalation, and other latent themes in legal intervention deaths of non-hispanic Black and White men: Results from the 2003--2017 National Violent Death Reporting System},
  author={Arseniev-Koehler, Alina and Foster, Jacob Gates and Mays, Vickie M and Chang, Kai-Wei and Cochran, Susan D},
  journal={American journal of public health},
  volume={111},
  number={S2},
  pages={S107--S115},
  year={2021},
  publisher={American Public Health Association}
}

@inproceedings{davidson2021exploring,
  title={Exploring nurse suicide by firearms: A mixed-method longitudinal (2003--2017) analysis of death investigations},
  author={Davidson, Judy E and Ye, Gordon and Deskins, Felicia and Rizzo, Heather and Moutier, Christine and Zisook, Sidney},
  booktitle={Nursing forum},
  volume={56},
  number={2},
  pages={264--272},
  year={2021},
  organization={Wiley Online Library}
}

@article{liu2023surveillance,
  title={Surveillance for violent deaths—national violent death reporting system, 48 states, the district of Columbia, and Puerto Rico, 2020},
  author={Liu, Grace S},
  journal={MMWR. Surveillance Summaries},
  volume={72},
  year={2023}
}

@article{ziems2024can,
  title={Can large language models transform computational social science?},
  author={Ziems, Caleb and Held, William and Shaikh, Omar and Chen, Jiaao and Zhang, Zhehao and Yang, Diyi},
  journal={Computational Linguistics},
  volume={50},
  number={1},
  pages={237--291},
  year={2024},
  publisher={MIT Press One Broadway, 12th Floor, Cambridge, Massachusetts 02142, USA~…}
}

@article{blair2016national,
  title={The national violent death reporting system: overview and future directions},
  author={Blair, Janet M and Fowler, Katherine A and Jack, Shane PD and Crosby, Alexander E},
  journal={Injury prevention},
  volume={22},
  number={Suppl 1},
  pages={i6--i11},
  year={2016},
  publisher={BMJ Publishing Group Ltd}
}

@inproceedings{ranjit-etal-2024-oath,
    title = "{OATH}-Frames: Characterizing Online Attitudes Towards Homelessness with {LLM} Assistants",
    author = "Ranjit, Jaspreet  and
      Joshi, Brihi  and
      Dorn, Rebecca  and
      Petry, Laura  and
      Koumoundouros, Olga  and
      Bottarini, Jayne  and
      Liu, Peichen  and
      Rice, Eric  and
      Swayamdipta, Swabha",
    editor = "Al-Onaizan, Yaser  and
      Bansal, Mohit  and
      Chen, Yun-Nung",
    booktitle = "Proceedings of the 2024 Conference on Empirical Methods in Natural Language Processing",
    month = nov,
    year = "2024",
    address = "Miami, Florida, USA",
    publisher = "Association for Computational Linguistics",
    url = "https://aclanthology.org/2024.emnlp-main.724/",
    doi = "10.18653/v1/2024.emnlp-main.724",
    pages = "13033--13059"
}

@article{douze2024faiss,
      title={The Faiss library},
      author={Matthijs Douze and Alexandr Guzhva and Chengqi Deng and Jeff Johnson and Gergely Szilvasy and Pierre-Emmanuel Mazaré and Maria Lomeli and Lucas Hosseini and Hervé Jégou},
      year={2024},
      eprint={2401.08281},
      archivePrefix={arXiv},
      primaryClass={cs.LG}
}

@article{steenkamp2006national,
  title={The National Violent Death Reporting System: an exciting new tool for public health surveillance},
  author={Steenkamp, Malinda and Frazier, Lorraine and Lipskiy, N and DeBerry, M and Thomas, S and Barker, L and Karch, Debra},
  journal={Injury prevention},
  volume={12},
  number={suppl 2},
  pages={ii3--ii5},
  year={2006},
  publisher={BMJ Publishing Group Ltd}
}

@article{krippendorff1970estimating,
  title={Estimating the reliability, systematic error and random error of interval data},
  author={Krippendorff, Klaus},
  journal={Educational and psychological measurement},
  volume={30},
  number={1},
  pages={61--70},
  year={1970},
  publisher={Sage Publications Sage CA: Thousand Oaks, CA}
}

@article{cammack2024vital,
  title={Vital Signs: Suicide Rates and Selected County-Level Factors—United States, 2022},
  author={Cammack, Alison L},
  journal={MMWR. Morbidity and Mortality Weekly Report},
  volume={73},
  year={2024}
}

@article{wilson2022surveillance,
  title={Surveillance for violent deaths—national violent death reporting system, 42 states, the District of Columbia, and Puerto Rico, 2019},
  author={Wilson, Rebecca F},
  journal={MMWR. Surveillance Summaries},
  volume={71},
  year={2022}
}

@article{fincham2008impact,
  title={The impact of working with disturbing secondary data: Reading suicide files in a coroner's office},
  author={Fincham, Ben and Scourfield, Jonathan and Langer, Susanne},
  journal={Qualitative Health Research},
  volume={18},
  number={6},
  pages={853--862},
  year={2008},
  publisher={Sage Publications Sage CA: Los Angeles, CA}
}

@book{efron1994introduction,
  title={An introduction to the bootstrap},
  author={Efron, Bradley and Tibshirani, Robert J},
  year={1994},
  publisher={Chapman and Hall/CRC}
}

@article{sinyor2024effect,
  title={The effect of economic downturn, financial hardship, unemployment, and relevant government responses on suicide},
  author={Sinyor, Mark and Silverman, Morton and Pirkis, Jane and Hawton, Keith},
  journal={The Lancet Public Health},
  volume={9},
  number={10},
  pages={e802--e806},
  year={2024},
  publisher={Elsevier}
}

@inproceedings{reimers-2019-sentence-bert,
  title = "Sentence-BERT: Sentence Embeddings using Siamese BERT-Networks",
  author = "Reimers, Nils and Gurevych, Iryna",
  booktitle = "Proceedings of the 2019 Conference on Empirical Methods in Natural Language Processing",
  month = "11",
  year = "2019",
  publisher = "Association for Computational Linguistics",
  url = "https://arxiv.org/abs/1908.10084",
}

@inproceedings{gupta-etal-2023-coverage,
    title = "Coverage-based Example Selection for In-Context Learning",
    author = "Gupta, Shivanshu  and
      Gardner, Matt  and
      Singh, Sameer",
    editor = "Bouamor, Houda  and
      Pino, Juan  and
      Bali, Kalika",
    booktitle = "Findings of the Association for Computational Linguistics: EMNLP 2023",
    month = dec,
    year = "2023",
    address = "Singapore",
    publisher = "Association for Computational Linguistics",
    url = "https://aclanthology.org/2023.findings-emnlp.930/",
    doi = "10.18653/v1/2023.findings-emnlp.930",
    pages = "13924--13950"
}

@article{paulozzi2004cdc,
  title={CDC’s national violent death reporting system: background and methodology},
  author={Paulozzi, Leonard J and Mercy, J and Frazier, Lorraine and Annest, J Lee},
  journal={Injury prevention},
  volume={10},
  number={1},
  pages={47--52},
  year={2004},
  publisher={BMJ Publishing Group Ltd}
}

@book{groundingtheory,
  title     = "Discovery of Grounded Theory Strategies for Qualitative Research",
  author    = "Barney G. Glaser and Anselm L. Strauss",
  year      = 1967,
  publisher = "AldineTransaction ",
  address   = "London"
}

@inproceedings{barany2024chatgpt,
  title={ChatGPT for education research: exploring the potential of large language models for qualitative codebook development},
  author={Barany, Amanda and Nasiar, Nidhi and Porter, Chelsea and Zambrano, Andres Felipe and Andres, Alexandra L and Bright, Dara and Shah, Mamta and Liu, Xiner and Gao, Sabrina and Zhang, Jiayi and others},
  booktitle={International conference on artificial intelligence in education},
  pages={134--149},
  year={2024},
  organization={Springer}
}

@article{katz2024thematic,
  title={Thematic Analysis with Open-Source Generative AI and Machine Learning: A New Method for Inductive Qualitative Codebook Development},
  author={Katz, Andrew and Fleming, Gabriella Coloyan and Main, Joyce},
  journal={arXiv preprint arXiv:2410.03721},
  year={2024}
}

@article{tai2024examination,
  title={An examination of the use of large language models to aid analysis of textual data},
  author={Tai, Robert H and Bentley, Lillian R and Xia, Xin and Sitt, Jason M and Fankhauser, Sarah C and Chicas-Mosier, Ana M and Monteith, Barnas G},
  journal={International Journal of Qualitative Methods},
  volume={23},
  pages={16094069241231168},
  year={2024},
  publisher={SAGE Publications Sage CA: Los Angeles, CA}
}

@article{dai2023llm,
  title={LLM-in-the-loop: Leveraging large language model for thematic analysis},
  author={Dai, Shih-Chieh and Xiong, Aiping and Ku, Lun-Wei},
  journal={arXiv preprint arXiv:2310.15100},
  year={2023}
}

@inproceedings{xiao2023supporting,
  title={Supporting qualitative analysis with large language models: Combining codebook with GPT-3 for deductive coding},
  author={Xiao, Ziang and Yuan, Xingdi and Liao, Q Vera and Abdelghani, Rania and Oudeyer, Pierre-Yves},
  booktitle={Companion proceedings of the 28th international conference on intelligent user interfaces},
  pages={75--78},
  year={2023}
}

@article{perkins2024use,
  title={The use of Generative AI in qualitative analysis: Inductive thematic analysis with ChatGPT},
  author={Perkins, Mike and Roe, Jasper},
  journal={Journal of Applied Learning and Teaching},
  volume={7},
  number={1},
  year={2024}
}

@inproceedings{blosnich2024industries,
  title={Industries of disruption: New avenues for upstream suicide prevention},
  author={Blosnich, John and Ward, Jeanne and Haydinger, Alexandra and Perkins, Melissa and De Luca, Susa},
  booktitle={APHA 2024 Annual Meeting and Expo},
  year={2024},
  organization={APHA}
}

@article{rockett2018discerning,
  title={Discerning suicide in drug intoxication deaths: Paucity and primacy of suicide notes and psychiatric history},
  author={Rockett, Ian RH and Caine, Eric D and Connery, Hilary S and D’Onofrio, Gail and Gunnell, David J and Miller, Ted R and Nolte, Kurt B and Kaplan, Mark S and Kapusta, Nestor D and Lilly, Christa L and others},
  journal={PLoS one},
  volume={13},
  number={1},
  pages={e0190200},
  year={2018},
  publisher={Public Library of Science San Francisco, CA USA}
}

@article{pestian2012sentiment,
  title={Sentiment analysis of suicide notes: A shared task},
  author={Pestian, John P and Matykiewicz, Pawel and Linn-Gust, Michelle and South, Brett and Uzuner, Ozlem and Wiebe, Jan and Cohen, K Bretonnel and Hurdle, John and Brew, Christopher},
  journal={Biomedical informatics insights},
  volume={5},
  pages={BII--S9042},
  year={2012},
  publisher={SAGE Publications Sage UK: London, England}
}

@misc{dasgupta2021ghost,
  title={Ghost in the machine: The emotional gravity of conducting mortality research},
  author={Dasgupta, Nabarun},
  journal={American Journal of Public Health},
  volume={111},
  number={S2},
  pages={S80--S81},
  year={2021},
  publisher={American Public Health Association}
}

@article{arseniev2023gendered,
  title={Gendered Patterns in Manifest and Latent Mental Health Indicators Among Suicide Decedents: 2003--2020 National Violent Death Reporting System (NVDRS)},
  author={Arseniev-Koehler, Alina and Mays, Vickie M and Foster, Jacob G and Chang, Kai-Wei and Cochran, Susan D},
  journal={American journal of public health},
  volume={114},
  number={S3},
  pages={S268--S277},
  year={2023},
  publisher={American Public Health Association}
}

@article{chance2025measuring,
  title={Measuring Narrative Complexity Among Suicide Deaths in the National Violent Death Reporting System (2003--2021 NVDRS)},
  author={Chance, Christina and Arseniev-Koehler, Alina and Mays, Vickie M and Chang, Kai-Wei and Cochran, Susan D},
  journal={Information},
  volume={16},
  number={11},
  pages={989},
  year={2025},
  publisher={MDPI}
}

@article{kirtley2022translating,
  title={Translating promise into practice: a review of machine learning in suicide research and prevention},
  author={Kirtley, Olivia J and van Mens, Kasper and Hoogendoorn, Mark and Kapur, Navneet and De Beurs, Derek},
  journal={The Lancet Psychiatry},
  volume={9},
  number={3},
  pages={243--252},
  year={2022},
  publisher={Elsevier}
}

@article{lejeune2022artificial,
  title={Artificial intelligence and suicide prevention: a systematic review},
  author={Lejeune, Alban and Le Glaz, Aziliz and Perron, Pierre-Antoine and Sebti, Johan and Baca-Garcia, Enrique and Walter, Michel and Lemey, Christophe and Berrouiguet, Sofian},
  journal={European psychiatry},
  volume={65},
  number={1},
  pages={e19},
  year={2022},
  publisher={Cambridge University Press}
}

@article{ehtemam2024role,
  title={Role of machine learning algorithms in suicide risk prediction: a systematic review-meta analysis of clinical studies},
  author={Ehtemam, Houriyeh and Sadeghi Esfahlani, Shabnam and Sanaei, Alireza and Ghaemi, Mohammad Mehdi and Hajesmaeel-Gohari, Sadrieh and Rahimisadegh, Rohaneh and Bahaadinbeigy, Kambiz and Ghasemian, Fahimeh and Shirvani, Hassan},
  journal={BMC medical informatics and decision making},
  volume={24},
  number={1},
  pages={138},
  year={2024},
  publisher={Springer}
}

@article{melia2025application,
  title={The application of AI to ecological momentary assessment data in suicide research: Systematic review},
  author={Melia, Ruth and Musacchio Schafer, Katherine and Rogers, Megan L and Wilson-Lemoine, Emma and Joiner, Thomas Ellis},
  journal={Journal of Medical Internet Research},
  volume={27},
  pages={e63192},
  year={2025},
  publisher={JMIR Publications Toronto, Canada}
}

@article{wang2024large,
  title={Large language models that replace human participants can harmfully misportray and flatten identity groups},
  author={Wang, Angelina and Morgenstern, Jamie and Dickerson, John P},
  journal={arXiv preprint arXiv:2402.01908},
  year={2024}
}

@article {Kafka125,
	author = {Kafka, Julie M and Moracco, Kathryn Elizabeth and Pence, Brian W and Trangenstein, Pamela J and Fliss, Mike Dolan and McNaughton Reyes, Luz},
	title = {Intimate partner violence and suicide mortality: a cross-sectional study using machine learning and natural language processing of suicide data from 43 states},
	volume = {30},
	number = {2},
	pages = {125--131},
	year = {2024},
	doi = {10.1136/ip-2023-044976},
	publisher = {BMJ Publishing Group Ltd},
	issn = {1353-8047},
	URL = {https://injuryprevention.bmj.com/content/30/2/125},
	eprint = {https://injuryprevention.bmj.com/content/30/2/125.full.pdf},
	journal = {Injury Prevention}
}

@article{parker2025supervised,
  title={Supervised Natural Language Processing Classification of Violent Death Narratives: Development and Assessment of a Compact Large Language Model},
  author={Parker, Susan T},
  journal={JMIR AI},
  volume={4},
  pages={e68212},
  year={2025},
  publisher={JMIR Publications Toronto, Canada}
}

@article{wang2025multi,
  title={A multi-stage large language model framework for extracting suicide-related social determinants of health},
  author={Wang, Song and Wei, Yishu and Ma, Haotian and Lovitt, Max and Deng, Kelly and Meng, Yuan and Xu, Zihan and Zhang, Jingze and Xiao, Yunyu and Ding, Ying and others},
  journal={Communications Medicine},
  volume={5},
  number={1},
  pages={404},
  year={2025},
  publisher={Nature Publishing Group UK London}
}

@article{parker2025assessing,
  title={Assessing Supervised Natural Language Processing (NLP) Classification of Violent Death Narratives: Development and Assessment of a Compact Large Language Model (LLM) Approach},
  author={Parker, Susan T},
  journal={medRxiv},
  pages={2025--01},
  year={2025},
  publisher={Cold Spring Harbor Laboratory Press}
}

@book{teutsch2000principles,
  title={Principles and practice of public health surveillance},
  author={Teutsch, Steven M and Churchill, R Elliott},
  year={2000},
  publisher={Oxford University Press}
}

\clearpage
\appendix

\section*{Appendix}

\section{Agreement Across Models}
\label{sec:app_agreement}

\begin{figure*}[h!]
  \includegraphics[width=0.95\textwidth]{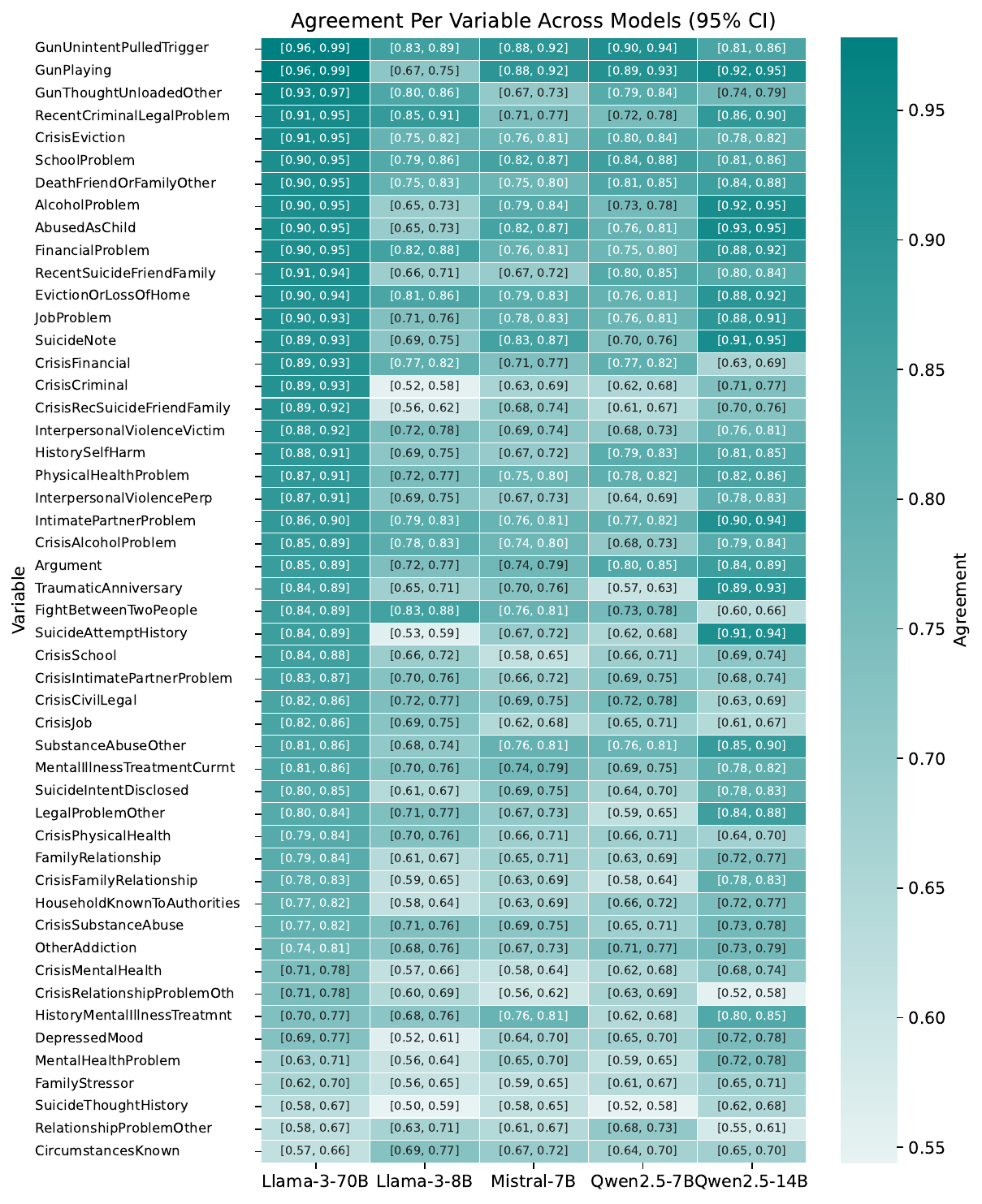}
  \caption{Per variable agreement (95\% confidence intervals; bootstrapped with 10K iterations) for  for 50 NVDRS variables across different models. Agreement is reported on $D_\text{balanced}$. We find that on average, \llama has the highest agreement with data annotators out of all evaluated models.}
  \label{fig:per_variable}
\end{figure*}

We use LMs as annotation assistants to label 36 circumstance and 14 crisis variables in NVDRS death narratives in \textsection \ref{sec:llm_annotations}. We evaluated LM agreement with data annotators on a test set: $D_\text{balanced}$ composed of 500 narratives per variable (with equal representation across 0/1 classes). The per variable agreement across models is shown in \autoref{fig:per_variable}. On average, \llama has the highest agreement with data annotators. However, there are a few outstanding cases where smaller models such as \texttt{Qwen2.5-14B} have higher agreement with the annotator (e.g. HistoryMentalIllnessTreatment). 

\subsection{Surfacing Annotation Inconsistencies with LMs}
\label{sec:app_inconsistencies}
The suicide prevention expert analyzed 300 individual examples sampled from D\_balanced across six variables where \llama either disagreed or agreed with the original human annotation. We calculated the overall proportion in the disagreement / agreement cases where the expert agreed with the model. 
Our bootstrap hypothesis test for equality of means accounts for sampling variability and shows a statistically significant difference ($p<0.05$) between the 38\% of cases where the model surfaced annotation inconsistencies on disagreement cases and the 13\% rate of joint human-model inconsistencies on agreement cases. We make the assumption that small shifts in model agreement (within the ~9\% CI range) are unlikely to shift the underlying error patterns substantially and would only affect the variability in the number of samples that constitute the distinct 13\% and 38\% conditional probabilities. While simulating robustness under lower model accuracy is possible (e.g., randomly flipping predictions to reduce agreement and directly recalculating ‘model corrects human’ and ‘joint error’ rates), this would introduce a new assumption that accuracy changes occur via random prediction flips rather than a more systematic variance that occurs for ambiguous or difficult samples. Therefore, we rely on the results of our bootstrap hypothesis test as a signal of robustness. 

\begin{table*}[t]
\centering
\small
\begin{tabular}{lcccc}
\toprule
\textbf{Variable} & \textbf{Agreement} & \textbf{TPR} & \textbf{FPR} & \textbf{FNR} \\
\midrule
GunUnintentPulledTrigger     & 0.978 & 0.984 & 0.028 & 0.016 \\
GunPlaying                   & 0.976 & 0.968 & 0.016 & 0.032 \\
GunThoughtUnloadedOther      & 0.954 & 0.992 & 0.084 & 0.008 \\
RecentCriminalLegalProblem   & 0.934 & 0.908 & 0.040 & 0.092 \\
CrisisEviction               & 0.932 & 0.924 & 0.060 & 0.076 \\
SchoolProblem                & 0.926 & 0.996 & 0.144 & 0.004 \\
HistoryMentalIllnessTreatmnt & 0.736 & 0.484 & 0.012 & 0.516 \\
DepressedMood                & 0.732 & 0.508 & 0.044 & 0.492 \\
MentalHealthProblem          & 0.672 & 0.352 & 0.008 & 0.648 \\
FamilyStressor               & 0.662 & 0.592 & 0.268 & 0.408 \\
SuicideThoughtHistory        & 0.628 & 0.268 & 0.012 & 0.732 \\
RelationshipProblemOther     & 0.624 & 0.964 & 0.716 & 0.036 \\
CircumstancesKnown           & 0.614 & 0.232 & 0.004 & 0.768 \\
\bottomrule
\end{tabular}
\caption{We report the agreement, true positive rate (TPR), false positive rate (FPR), and false negative rate (FNR) for a subset of variables with highest and lowest agreements. 
Performance is reported on the balanced evaluation set ($D_\text{balanced}$) using \llama.}
\label{tab:tpr_fnr}
\end{table*}

\begin{table}[t]
\centering
\small
\begin{tabular}{lcc}
\toprule
\textbf{Model} & \textbf{Mean Agreement} & \textbf{S.D. across vars.} \\
\midrule
\textbf{Llama-3-70B} & \textbf{0.82, [0.79, 0.85]} & \textbf{0.12}\\
Qwen2.5-14B & 0.81, [0.80, 0.82] & 0.09 \\
Qwen2.5-7B & 0.56, [0.54, 0.58] & 0.17\\
Mistral-7B & 0.67, [0.65, 0.69] & 0.14 \\
Llama-3-8B & 0.54, [0.52, 0.56] & 0.17 \\
\bottomrule
\end{tabular}
\caption{Mean agreement, 95\% confidence intervals, and standard deviation across 50 variables for different models. \llama achieves the highest agreement of 82\% with data annotators. 
Performance is reported on a random evaluation set of 1000 narratives with unequal representation across 0/1 classes ($D_\text{random}$).}
\label{tab:random_eval}
\end{table}



\begin{table}[h!]
\centering
\scalebox{0.75}{
\begin{tabular}{cc}
\toprule
\textbf{Config}&\textbf{Assignment}\\
\midrule
models &
\makecell[c]{
\textbf{Meta-Llama-3-70B-Instruct}\\
Number of parameters: 70B
\\
\midrule
\textbf{Meta-Llama-3-8B-Instruct}\\
Number of parameters: 8B
\\
\midrule
\textbf{Mistral-7B-Instruct-v0.2}\\
Number of parameters: 7B
\\
\midrule
\textbf{Qwen2.5-14B-Instruct-1M}\\
Number of parameters: 14B\\
\midrule
\textbf{Qwen2.5-7B-Instruct-1M}\\
Number of parameters: 7B\\
\midrule
}\\
GPU& NVIDIA RTX A6000\\
\# of GPUs & 4 (inference)\\
\bottomrule
\end{tabular}
}
\caption{We provide details about the models used for our experiments in \textsection \ref{sec:llm_validation} and \textsection \ref{sec:agent-in-the-loop}}
\label{tab:model_list}
\end{table}

\section{LM-Simulated Codebook Development}
\label{sec:simulated_codebook_development_details}

The expert feedback $e$ we use in the simulated setting is the LM CoT, and in order to avoid faulty $e$, we only keep the samples for which the LM predictions matched the abstractor labels. 
This can be considered a disadvantage to the codebook that gets generated from our codebook development algorithm compared to the reference codebook because the algorithm is limited to receiving feedback on potentially easy cases for which the LM was already able to predict correctly using the reference codebook. 
However, despite this disadvantage, the generated codebook outperforms the reference codebook as shown in the results in \autoref{fig:nvdrs_hitl_acc_sim}. 
We suspect this stronger performance to be attributed to our algorithm taking a systematic approach to refining the codebook based on individual examples and therefore providing better granular instructions until a performance threshold is reached, as opposed to the reference codebook which is developed more holistically and therefore overlook details.

\begin{figure*}[h!]
  \includegraphics[width=\textwidth]{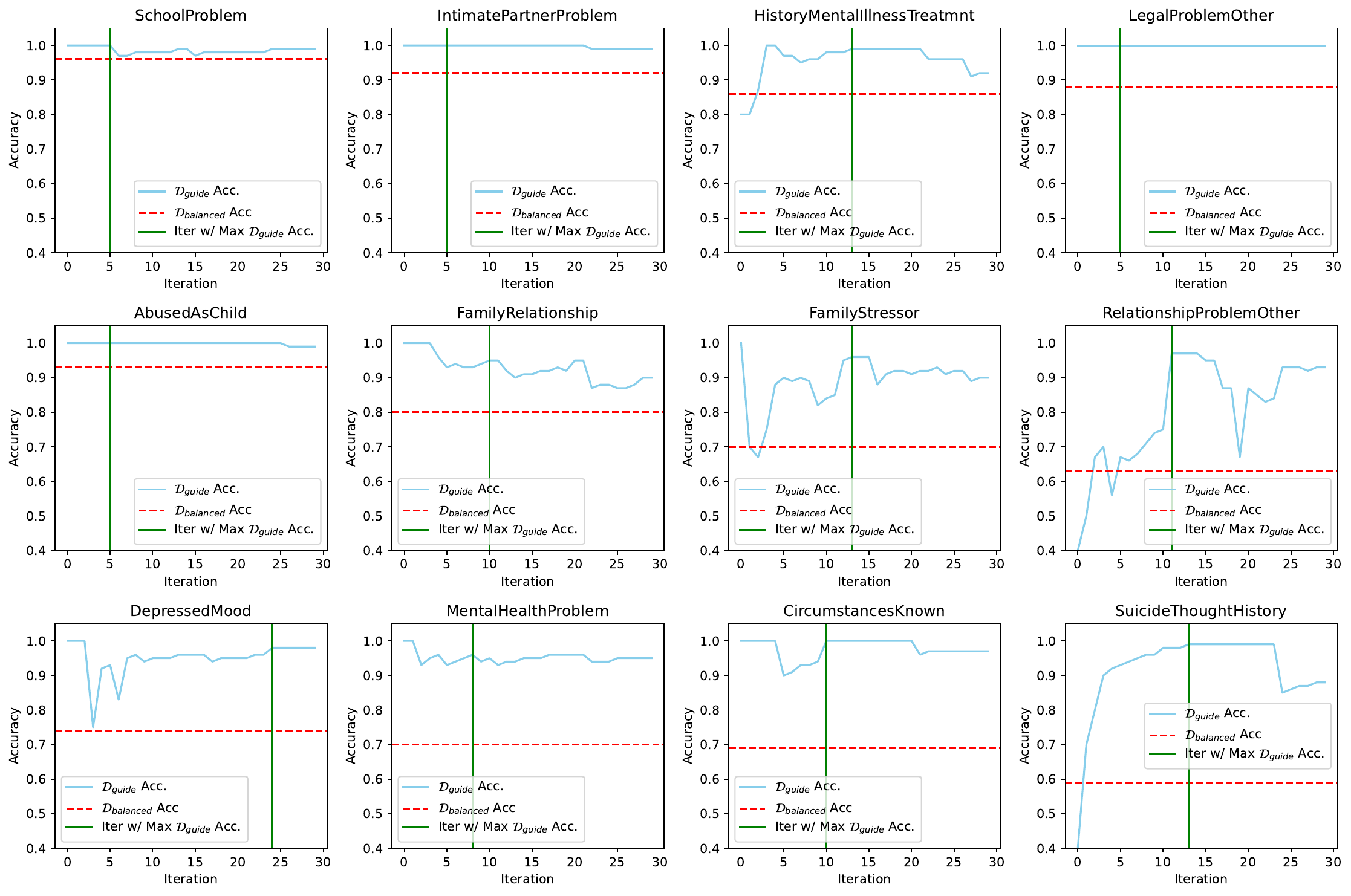}
  \caption{Accuracy on $\mathcal{D}_{guide}$ across 30 iterations for LM-Simulated codebook development for 12 NVDRS variables. The maximum accuracy on $\mathcal{D}_{guide}$ is reached between 10-15 iterations for all variables using random sampling per iteration.
  }
  \label{fig:llm_sim_all_var}
\end{figure*}

\autoref{fig:llm_sim_all_var} shows the accuracy on $\mathcal{D}_{guide}$ across 30 iterations. Most of the variables reach the max accuracy between iterations 10-15. Furthermore, we see greater instability in performance in earlier iterations due to the small size of $\mathcal{D}_{guide}$. 

\section{Coverage-based Sampling}
\label{sec:coverage-based-sampling}

Coverage is defined as how much of a sample's content overlaps in content with another set of samples. 
Coverage-based sampling is inspired by \cite{gupta-etal-2023-coverage}, which showed that selecting a set of samples by collective coverage leads to better performance than naively collecting samples by individual similarity. 
The main difference with our sampling strategy with \cite{gupta-etal-2023-coverage} is that instead of measuring coverage at the token level, we compute coverage at the sentence level of the retrieved sample. 
Given a set of narratives, each narrative is split into sentences, which are then embedded with the \texttt{all-MiniLM-L6-v2} model from SentenceTransformers\footnote{\url{https://sbert.net/index.html}}\citep{reimers-2019-sentence-bert}. 
The coverage of each sentence in a new sample is then computed by the maximum cosine similarity between the sentence and all other sentences in the set of chosen narratives. 
The coverage of the entire narrative is then the average value of these similarities. 
With all the coverage values computed for the set of samples to retrieve from, we select $N$ samples with least coverage to promote diversity. 

\section{Codebook Development Algorithm Hyperparameters}
\label{sec:app_hyperparameters}
\autoref{tab:llm_hitl_config} provides an overview of hyperparameters for the codebook development algorithm. $t^*$ is the number of iterations that the codebook development algorithm ran for. For all experiments, $t$ was fixed to 30. In practice, $t$ would vary depending on the performance $m$ on $\mathcal{D}_{guide}$. $b$ is the budget—the maximum number of narratives that the algorithm is allowed to iterate over. $n$ is the batch size per iteration. $n$ can be sampled using random or coverage-based sampling. 
Model\_id is the model used for codebook development in our algorithm. $k$ is the minimum size of $\mathcal{D}_{guide}$, and $m$ is the target performance for $\mathcal{D}_{guide}$. 
We leave it to expert judgment to determine $b$, $k$, and $m$, as these depend on the nature of the variable and thus, the number of iterations required to reach thematic saturation \citep{groundingtheory}. Experts may consider the number of consecutive iterations without updating the codebook as an additional hyperparameter for determining the stopping condition.
For our legal interaction case study, we set $b$ as 150 given experts analyzed 150 narratives manually in one round of qualitative coding to develop the codebook manually. 
We set $k$ by observing performance trends on $\mathcal{D}_{guide}$ in the LM-simulated codebook development experiments (\textsection \ref{sec:agent-in-the-loop} where performance was unstable in the first 5 iterations. 
 
\begin{table*}[ht]
\centering
\small
\begin{tabular}{lccccccc}  
\toprule
\textbf{Model} & \textbf{$t$*} & \textbf{$b$} & \textbf{$n$} & \textbf{sampling} & \textbf{model\_id} & \textbf{$k$} & \textbf{$m$} \\
\midrule
LM-Sim (NVDRS) & 30 & 150 & 5 & Random/Coverage & \llama & - & -\\  
\hitl (Legal) & 30 & 150 & 5 & Coverage & \llama & 30 & 0.9 \\  
\bottomrule
\end{tabular}
\caption{Configuration details for LM-Sim and \hitl across various parameters. }
\label{tab:llm_hitl_config}
\end{table*}

\begin{figure*}[h!]
  \includegraphics[width=\textwidth]{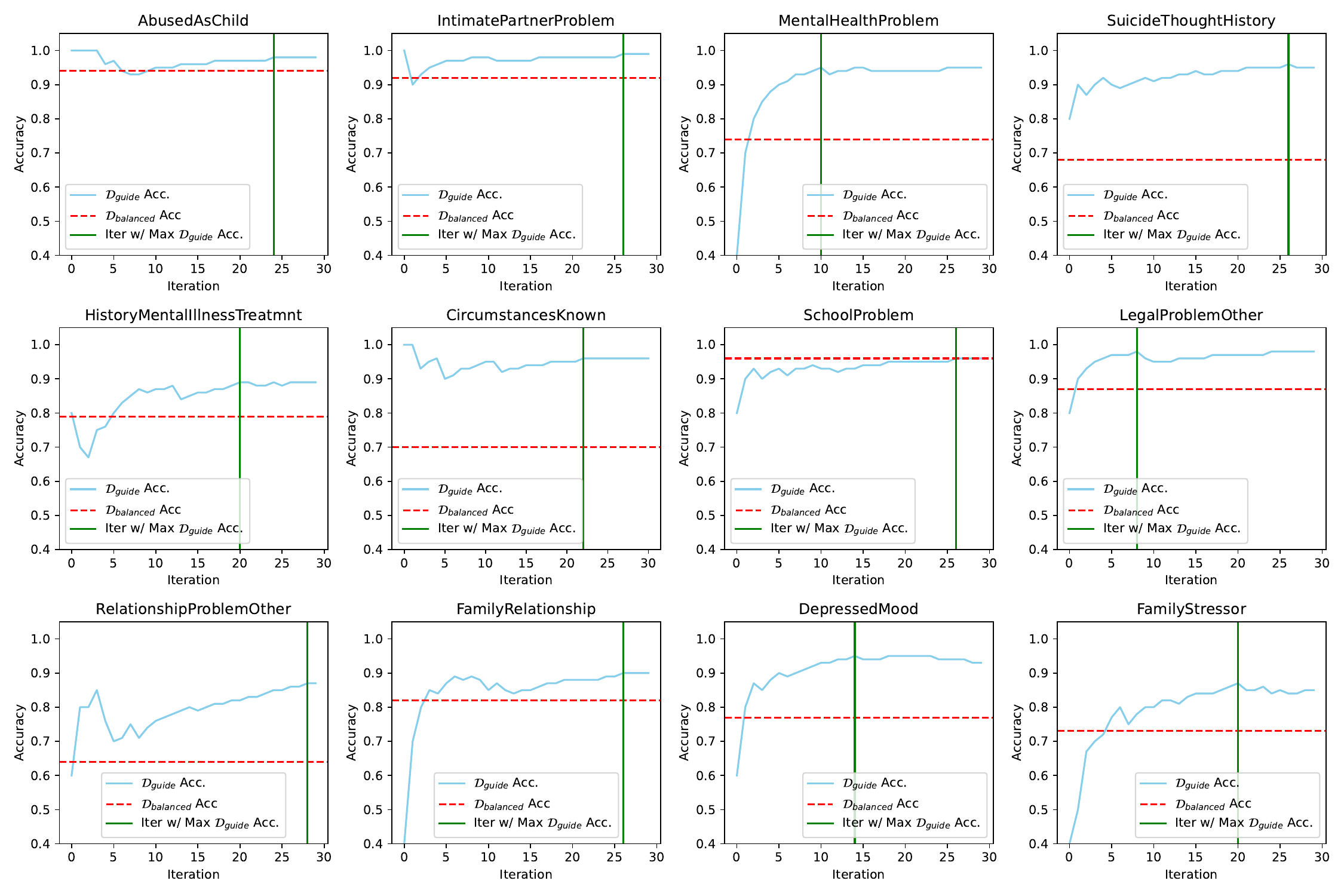}
  \caption{Accuracy on $\mathcal{D}_{guide}$ across 30 iterations for LM-Simulated codebook development for 12 NVDRS variables using coverage based sampling per iteration. 
  }
  \label{fig:llm_sim_all_var_coverage}
\end{figure*}

\section{Case Study: Legal Interactions}
\label{sec:app_case_study}

 In \autoref{tab:llm_hitl_config}, we show the hyperparameter configurations for both our LM-simulated codebook development and \hitl setting for legal interactions. In \autoref{tab:prompt_templates}, we show the initial prompt templates $\mathcal{G}_{\text{0}}$, for both the simulated and \hitl setting. \autoref{tab:hitl_expert_lawyer} shows our manually developed expert codebooks (left) and our \hitl codebooks (right). 

\begin{figure*}[h!]
  \includegraphics[width=\textwidth]{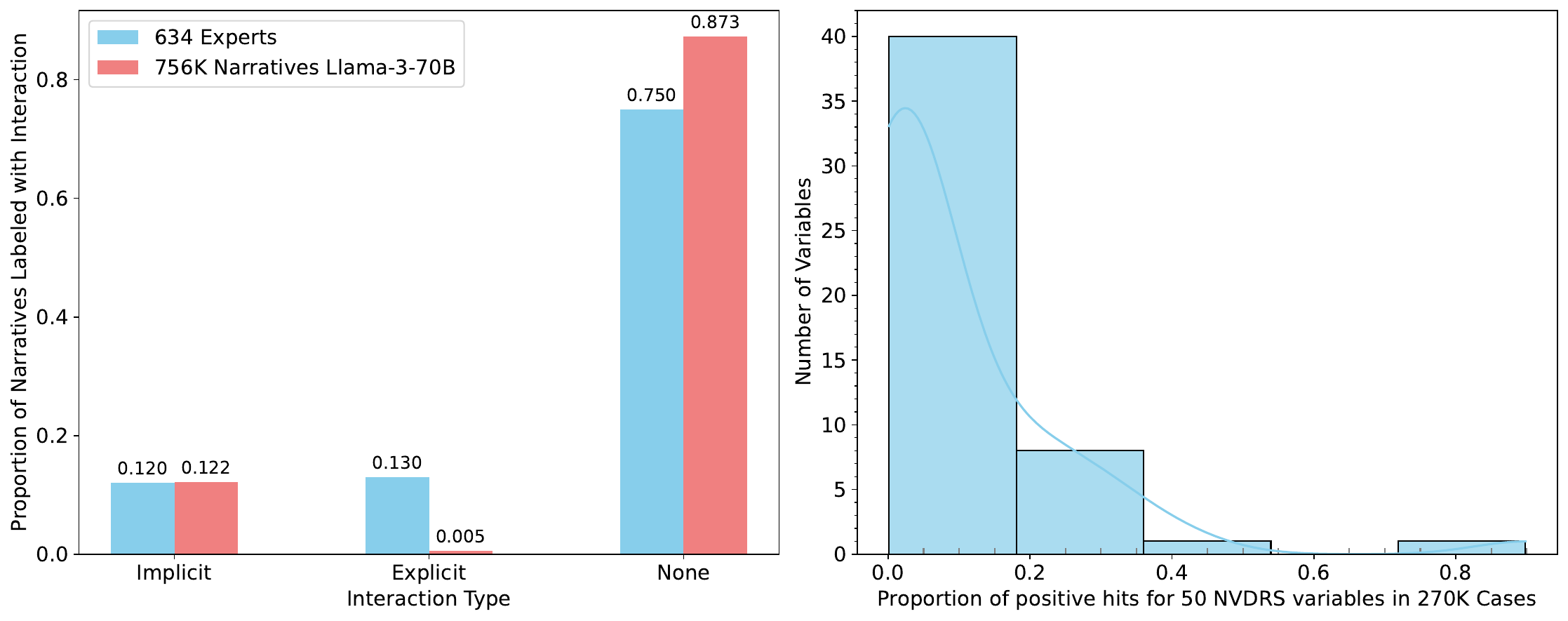}
  \caption{Distribution of narratives containing implicit-, explicit- and no-interaction for 3 data splits - 634 experts ($D_{\text{expert\_legal}}$), 270K CME narratives, and 130K LE narratives (left). Distribution of proportion of positive occurrences for 50 variables in 270K NVDRS cases (right).} 

  \label{fig:interaction_distribution}
\end{figure*}

\autoref{fig:interaction_distribution} (left) shows the distribution of narratives with implicit-, explicit-, and no interactions across three data splits. 
In \autoref{fig:interaction_distribution} (right), we show the distribution of the proportion of positive occurrences for 50 NVDRS variables in all 270K cases. This distribution has a heavy right skew showing the heavy class imbalance in NVDRS. 

\begin{table}[t]
\centering
\small
\begin{tabular}{lccc}
\toprule
\textbf{Annotated Set} & \textbf{Implicit} & \textbf{Explicit} & \textbf{None} \\
\midrule
$\mathcal{D}_{guide}$ & 55 & 23 & 72 \\
$\mathcal{D}_{val}$ & 20 & 20 & 20 \\
$D_{\text{expert\_legal}}$ & 74 & 83& 477\\
\bottomrule
\end{tabular}
\caption{Distribution across implicit, explicit and no-interactions for all annotated sets in legal interaction case study.}
\label{tab:appendix:casestudy_distributions}
\end{table}

We plot the F$_1$ on $\mathcal{D}_{guide}^{t}$ and $\mathcal{D}_{val}$ across iterations for all three interaction types in \autoref{fig:hitl_lawyer_iter}.\footnote{Please see \autoref{tab:appendix:casestudy_distributions} in Appendix \ref{sec:app_case_study} for $\mathcal{D}_{guide}^{30}$, $\mathcal{D}_{val}$ ($j$=20) and $D_{\text{expert\_legal}}$ class distributions.} 
As expected, the performance is unstable in the first few iterations given that $\mathcal{D}_{guide}^{t}$ only contains a few narratives. However, we observe that peak performance is reached by the 25th iteration (Macro F$_1$ on $\mathcal{D}_{val}$ was 0.8).\footnote{Please see \autoref{tab:llm_hitl_config} in Appendix \ref{sec:app_hyperparameters} for further discussion on hyperparameter selection for our case study.}
We hypothesize that performance on explicit interactions is less stable because most of the guidelines in the codebook pertain to implicit interactions, as shown in \autoref{tab:hitl_expert_lawyer}, resulting in limited instruction to identify explicit interactions. 

\begin{table*}[h!]
\centering
\scalebox{0.6}{
\begin{tabular}{|p{11cm}|p{11cm}|} 
\hline
\textbf{LM Simulation (NVDRS)} & \textbf{\hitl (Legal Interactions)} \\ 
\hline

\textbf{Prompt}

{\ttfamily
Instructions: You are an expert suicide caseworker and your job is to annotate reports with the \{code\_name\} variable. Do not read into the text and stick to the definition of variable strictly. If two reports are provided, use both reports to determine your response but only return one response for both reports with no additional text!
Provide the reasoning for your answer, the span of text that you used to generate your answer and your response using the response options only and return your answer in the following format: \{'reason': 'reasoning', 'span': 'span of text', 'response': '1.0 or 0.0'\}

}

\textbf{Guideline Synthesis Prompt}:  
{\ttfamily
You are an expert suicide caseworker and your job is to curate a set of guidelines that will be used by another model to label suicide reports with the variable:\{code\_name\}. You will be shown the original set of guidelines, the report that was used to label the variable \{code\_name\}, the model's label, the correct human label, the human's reasoning, and the span of text that the human used from reports to decide their label. The label can be 0.0 or 1.0. You have to return a set of new guidelines using this information which will be used to annotate \{code\_name\} for future reports. Keep the guidelines concise, and use the human reasoning, span, or other information from the report to update the guidelines, make sure to not lose out on information in the original set of guidelines but try not to have too much repetition. You have to return your answer in the following format with absolutely not additional text!: 'Guidelines: *..., *...'.
}

& 
\textbf{Prompt}:  
{\ttfamily

Instructions: You are an expert suicide caseworker trained to correctly categorize suicide reports by the victim's interaction with a lawyer or attorney. 
            You have to label each report with only one of the 3 interaction types and return your answer in the following format: \{reason: 'reasoning', span: 'span of text', label: 'implicit\_interaction, explicit\_interaction, no\_interaction'\}"
, with the reason behind your answer, the span of text you used to determine your answer, and a label and no additional text.
            If two reports are given, only return one answer using both reports using the format and make sure to provide which report you got the span from!

            Classes: 
        
            Label: no\_interaction
            \textbullet{} Definition: It is not implied or explicitly stated that V had interactions with a lawyer. 
            
            Label: implicit\_interaction
            \textbullet{} Definition: V had an implicit interaction with a lawyer where it is implied that V had an interaction with a lawyer. 

            Label: explicit\_interaction 
            \textbullet{}Definition: There are explicit mentions of V interacting with a lawyer or attorney.     

}  \\
\hline 
\end{tabular}
}
\caption{Prompt templates for LM simulated setting (NVDRS variables) and for \hitl codebook development for legal interactions. }
\label{tab:prompt_templates}
\end{table*}

\begin{table*}[h!]
\centering
\scalebox{0.6}{
\begin{tabular}{|p{11cm}|p{11cm}|} 
\hline
\textbf{Expert (Manual)} & \textbf{\hitl} \\ 
\hline
\textbf{Guidelines}:  
{\ttfamily
\textbullet{} Label:  no\_interaction  
Definition: It is not implied or explicitly stated that V had interactions with a lawyer. Being released from jail, arrest warrants, or being under investigation for a crime should be labeled with  no\_interaction  

Positive Example and Justification: `Censored': Being released from jail does not imply any interaction with a legal professional  
         
\textbullet{} Label: implicit\_interaction  

Definition: The report mentions circumstances such as divorce, separation, or issues surrounding child custody/visitation, and should be labeled with  implicit\_interaction . Additionally, mentions of court proceedings/appearance, court orderings, restraining orders, financial crimes, lawsuits or if V was charged/accused with severe crimes such as DUI/DWI, assaulting an officer, battery, domestic violence/protection orders, ongoing legal problems and arrests for severe crimes within the last 6 months etc. all imply interactions all should be labeled with  implicit\_interaction . 

Positive Example and Justification: `Censored': V was going through a divorce so it is implied they had an interaction with an attorney
            
\textbullet{} Label:  explicit\_interaction  

Definition: There are explicit mentions of V interacting with a lawyer or attorney. Only choose this label if the legal professional (i.e. lawyer) is directly mentioned in the report.

Positive Example and Justification: `Censored': It is explicitly stated that V missed appointments with his lawyer, so this is  explicit\_interaction  
} 
& 
\textbf{Guidelines}:  
{\ttfamily
\textbullet{}Litigation was noted as pending meaning it was scheduled for some future date, therefore it is unclear if victim had actually spoken with a lawyer at the time of death,  and this should be labeled as no interaction. 

\textbullet{}Typically if the victim, themselves, was an attorney, this should be labeled as no interaction. However, when the victim was an attorney and the victim had evidence of legal problems requiring some hearing, court interaction, or need of lawyer service, then this should be labeled as implicit if it is not stated that they did not directly interact with another lawyer or attorney. If they did interact with another lawyer or attorney, then this should be labeled as explicit. 

\textbullet{}Although the victim has bankruptcy paperwork, it is unclear if this paperwork was filed thus it is unclear if a lawyer or attorney was currently involved, and this should be labeled as no interaction. 

\textbullet{}Because the victim was facing criminal charges, this likely means a lawyer or attorney was involved in this legal proceeding at the time of death, and this should be labeled with implicit interaction. 

\textbullet{}Although the victim’s sale of his business was not going well, that phrase cannot be interpreted as indicating an implicit interaction with a lawyer or attorney, and this should be labeled as no interaction. 

\textbullet{}although the narrative mentions the victim had a nasty divorce, which would typically be an implicit interaction, it was noted the divorce was 2 years in the past, which means any current interactions with a lawyer or attorney is unlikely and this should be labeled as no interaction. 

\textbullet{}In this instance, the mention of lawyer or attorney is in reference to the sister of the victim and not the victim, themselves, and the sister talking to a lawyer seems to have been after the victim's death.  The victim, themselves, should be the one who had the interaction, or the family member who talked with a lawyer or attorney should have done so before the victim's death, then this should be labeled as implicit or explicit depending on whether the lawyer or attorney is noted. 

\textbullet{}In this narrative the IRS issues were framed as they were going to visit, which means this had not happened yet, therefore it is unclear if a lawyer or attorney was yet involved, and this should be labeled as no interaction. 

\textbullet{}If the victim is in the process of a divorce or if a divorce hearing is pending, then that should be labeled as implicit interaction. 

\textbullet{}The victim was considering bankruptcy, which means we do not know if a lawyer or attorney was involved, and this should be labeled as no interaction. 

\textbullet{}There was an ongoing custody issue that included a guardian ad litem, which is a court-appointed representative, so this should be labeled as implicit interaction. 

\textbullet{}The victim was facing jail time or imprisonment, and this should be labeled as implicit interaction. 

\textbullet{}Just because the victim was a law student, does not mean there was an interaction and should be labeled as no interaction.  

\textbullet{}The threat of being sued was not sufficient to imply a lawyer or attorney interaction and should be labeled as no interaction. 

\textbullet{}Because the narrative explains that victim was awaiting trial, that should be labeled as implicit. 

\textbullet{}Just because a complaint had been filed, that is not sufficient to assume a lawyer or attorney interaction, so this should be labeled as no interaction. 

[truncated ...]










}  \\
\hline 
\end{tabular}
}
\caption{Expert codebook (left) for defining legal interactions and \hitl codebook developed with suicide prevention expert (right).}
\label{tab:hitl_expert_lawyer}
\end{table*}

\section{NVDRS Codebooks vs LM Simulated Codebooks}
\label{sec:app_codebooks_comp}

We provide our generated codebooks (right) for 3 NVDRS variables. Codebooks generated with our LM-simulated pipeline contain finer-grained instruction and examples from narratives which could be helpful in the future for augmenting existing NVDRS codebooks. 
\begin{table*}[h!]
\centering
\scalebox{0.6}
{
\begin{tabular}{|p{3cm}|p{9cm}|p{10cm}|} 
\hline
\textbf{Variable} &
\textbf{NVDRS Codebook} & \textbf{LL-Simulated Codebook)} \\ 
\hline
\textbf{AbusedasChild}
& 
\textbf{Prompt}: 

{\ttfamily
The victim had a history of abuse (physical, sexual, or psychological) or neglect (physical, including
medical/dental, emotional, or educational neglect; or exposure to violent environments; or inadequate
supervision) as a child.

This variable more broadly captures victim’s experiences of abuse and neglect irrespective of its
relationship to the violent death. Code “Yes” if the victim experienced abuse or neglect, but there is no
direct link to the violent death, or the link is unknown.
\textbullet{}Do NOT code if the abuse or neglect directly causes or precipitated the death, instead code
abuse/neglect led to death.
\textbullet{} Code as “Yes” if the victim had been the victim of child abuse at any point in the past, even if the
victim is currently an adult.
\textbullet{}Code “Yes” if the evidence of ongoing abuse is suspected but not confirmed. Code “Yes” if autopsy
evidence reported an indication of previous abuse.
\textbullet{}Abuse can be physical, psychological, sexual, or other as long as the source document refers to
“abuse.” Neglect may be medical, physical, or emotional.
\textbullet{}Indicate the nature of the abuse or neglect in the incident narrative.
\textbullet{} Abuse could have been perpetrated by a caregiver or other individual.

}

& 
\textbf{Prompt}: 
{\ttfamily 
\textbullet{} label abusedaschild as 1.0 if the report explicitly mentions abuse or neglect in the victim's childhood, 
\textbullet{}label abusedaschild as 1.0 if the report suggests a potentially abusive relationship in adulthood, such as a history of domestic calls or restraining orders, as this may indicate a pattern of abuse, 
\textbullet{} label abusedaschild as 0.0 if there is no mention or indication of abuse or neglect in the victim's childhood or adulthood, 
\textbullet{}consider the entire report, including the victim's history and circumstances surrounding the death, but do not make inferences about childhood experiences based on adult behaviors or circumstances unless there is a clear pattern of abuse, 
\textbullet{}do not assume childhood abuse based on domestic disputes or arguments in adulthood, 
\textbullet{}absence of any mention of childhood abuse or neglect in the report indicates a label of 0.0.

}

  \\
\hline
\textbf{FamilyRelationship}
& 
\textbf{Prompt}

{\ttfamily

 \textbullet{} Code as 1.0 if at the time of the incident the victim was experiencing a relationship problem with a
family member other than an intimate partner (e.g., a child, mother, in-law), and this appears to have
contributed to the death. 

Suicide example: The victim is despondent over his argument with his parents and dies by suicide.

 \textbullet{}If the report indicates that the victim was “having relationship problems,” these should be
assumed to be intimate partner problems, and not problems with other friends or family.

 \textbullet{}This code should not be endorsed for custody disputes when the victim is a child because the relationship problem in these instances is typically not with a child or other non-intimate partner family member. If the custody dispute affects the relationship of the parent and child, then this should be coded “1.0.”

 \textbullet{} Narrative should contain an explanation of the relationship problem and identify the family
member with whom the victim had a problem.

 \textbullet{}This variable does NOT capture problems related to the family environment that do not explicitly pertain to the victim’s relationship with a family member (e.g., the stress of caring for a sick family member).

}
& 
\textbf{Prompt}: 

{\ttfamily

 \textbullet{} Label as 1.0 if the report mentions a conflict, argument, or strained relationship with a family member (e.g. parent, sibling, child) that contributed to the death, including instances where the family member's death anniversary is mentioned 

 \textbullet{}Label as 1.0 if the report mentions a problematic relationship with a family member that occurred shortly before the incident
 
 \textbullet{}Label as 1.0 if the report mentions a loss of custody or a sense of failure towards family members especially children, that contributed to the death

 \textbullet{}Label as 1.0 if the report mentions the victim being locked out of their residence by family members or being isolated from them, indicating a strained relationship
 
 \textbullet{} Label as 0.0 if the report does not mention a strained or problematic relationship with a family member, or if the family member is only mentioned as a discoverer of the body or as someone who had a normal conversation with the deceased prior to the death, or if the family member is mentioned in the context of a past traumatic event that is not directly related to a strained relationship
 
 \textbullet{} Consider the context of the report, if the victim's despondency and irrational behavior are related to other factors such as terminal illness, alcoholism, etc., and not directly related to a family member, label as 0.0.

}

  \\
\hline
\textbf{LegalProblemOther}
& 
\textbf{Prompt}
{\ttfamily
Civil legal (non-criminal) problem(s) appear to have contributed to the death
Code as “Yes,” if at the time of the incident the victim was facing civil legal problems, such as a divorce,
custody dispute, civil lawsuit, or legal problems that were unspecified as either criminal or civil, and
these problems appear to have contributed to the death.

}
& 
\textbf{Prompt}: 

{\ttfamily
\textbullet{} legalproblemother refers to any legal problem that is not related to criminal justice or immigration, such as civil legal problems like divorce, child custody, eviction notices, or upcoming court appearances, 
\textbullet{}a legal problem is considered "other" if it is not related to criminal justice or immigration, even if it is not explicitly mentioned as a contributing factor to the suicide, 
\textbullet{}the presence of a civil legal problem can be inferred from the context of the report, but should be distinguished from financial concerns or medical issues, 
\textbullet{}specifically look for mentions of court appearances, legal proceedings, or legal issues that are not related to criminal justice or immigration, 
\textbullet{} eviction notices or other civil legal problems that contribute to feelings of depression or hopelessness should be labeled as 1.0, 
\textbullet{} if there is no indication of a civil legal problem in the report, and the report only mentions financial or medical issues, label as 0.0.
}

 \\
\hline
\end{tabular}
}
\caption{NVRDS codebook guidelines for 3 variables (left) compared to codebooks generated in the LM simulated setting (right) in \textsection \ref{sec:agent-in-the-loop}}
\label{tab:codebook_comps}
\end{table*}

\begin{figure*}[h!]
  \includegraphics[width=\textwidth]{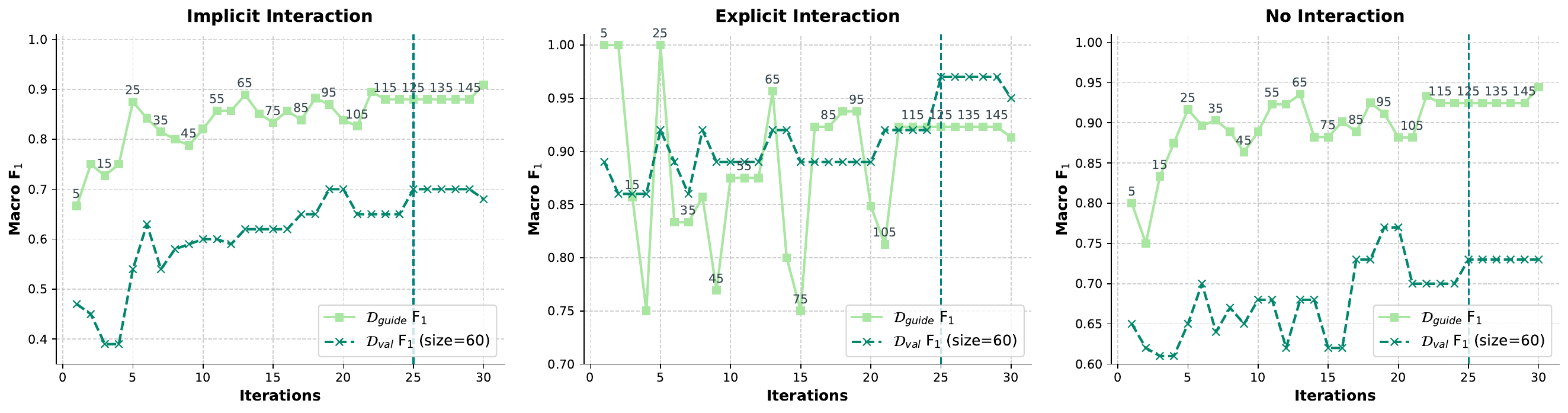}
  \caption{\llama performance on $\mathcal{D}_{val}$ and $\mathcal{D}_{guide}^{t}$ across 30 iterations for human-in-the-loop codebook development for detecting victim-lawyer interactions. 
  Data labels for $\mathcal{D}_{guide}$ represent the cumulative size of $\mathcal{D}_{guide}^{t}$ at each iteration $t$. 
  Max overall macro F$_1$ on $\mathcal{D}_{val}$ is reached by the 25th iteration (Macro F$_1$ of 0.8).
  }
  \label{fig:hitl_lawyer_iter}
\end{figure*} 
\section{Further Related Work}
\label{sec:app_relatedwork}
Prior work has explored the use of language models~\citep{parker2025assessing} to extract a range of suicide-related circumstances from unstructured narratives, including infrequent or highly specific variables such as circumstances preceding female firearm suicide~\cite{zhou2023identifying} and intimate partner violence~\cite{kafka2023detecting}.
Alternatively, \citet{wang2023nlp} finetune a BERT model to classify circumstance and crisis variables from narratives, while \citet{zhou2023identifying} use language models to identify infrequent circumstances preceding female firearm suicide using a manually developed codebook. 
\citet{kafka2023detecting} applies supervised learning to detect intimate partner violence (IPV), but found that their approach does not capture implicit references due to long narrative lengths.
For example, \citet{consoli2024sdoh} and \citet{guevara2024large} use LMs with in-context learning to identify SDoH in electronic health record (EHR) and medical notes. 

These efforts largely treat existing annotations and codebooks as fixed, and do not quantify how model performance varies across variables or characterize corresponding failure modes.
More importantly, prior work typically frames LMs as annotation assistants, without enabling the discovery of \textit{new} intervention opportunities. 
Motivated by these limitations, we examine whether LMs can serve as effective assistants to (i) data annotators by peer-validating existing annotations and surfacing potential discrepancies, and (ii) experts by supporting the development of codebooks for annotating new variables, surfacing new intervention opportunities not captured in the structured data.
\end{document}